\documentclass[10pt]{article}

\usepackage{amsmath}
\usepackage{amssymb}

\usepackage{graphicx}

\usepackage{cite,url}

\usepackage{color}

\usepackage[labelfont=bf,labelsep=period,justification=raggedright]{caption}

\makeatletter
\renewcommand{\@biblabel}[1]{\quad#1.}
\makeatother

\date{}

\begin{document}

\begin{flushleft}
{\Large
\textbf{Strategies used as spectroscopy of financial markets reveal new stylized facts}
}
\\
Wei-Xing Zhou$^{1,2,3,\ast}$,
Guo-Hua Mu$^{1,2,3}$,
Wei Chen$^{4}$,
Didier Sornette$^{5,6,\ast}$
\\
\bf{1} School of Business, East China University of Science and Technology, Shanghai, China
\\
\bf{2} School of Science, East China University of Science and Technology, Shanghai, China
\\
\bf{3} Research Center for Econophysics, East China University of Science and Technology, Shanghai, China
\\
\bf{4} Shenzhen Stock Exchange, 5045 Shennan East Road, Shenzhen 518010, China
\\
\bf{5} Department of Management, Technology and Economics, ETH Zurich, Zurich, Switzerland
\\
\bf{6} Swiss Finance Institute, c/o University of Geneva, Geneva, Switzerland
\\
$\ast$ E-mail: wxzhou@ecust.edu.cn (WXZ); dsornette@ethz.ch (DS)
\end{flushleft}

\section*{Abstract}
We propose a new set of stylized facts quantifying the structure of financial markets. The key idea is to study the combined structure of both investment strategies and prices in order to open a qualitatively new level of understanding of financial and economic markets. We study the detailed order flow on the Shenzhen Stock Exchange of China for the whole  year of 2003. This enormous dataset allows us to compare (i) a closed national market (A-shares) with an international market (B-shares), (ii) individuals and institutions and (iii) real investors to random strategies with respect to timing that share otherwise all other characteristics. We find that more trading results in smaller net return due to trading frictions. We unveiled quantitative power laws with non-trivial exponents, that quantify the deterioration of performance with frequency and with holding period of the strategies used by investors. Random strategies are found to perform much better than real ones, both for winners and losers. Surprising large arbitrage opportunities exist, especially when using zero-intelligence strategies. This is a diagnostic of possible inefficiencies of these financial markets.


\section*{Introduction}

\label{S1:intro}

Nothing in biology makes sense except in the light of evolution. This famous sentence by Theodosius Dobzhanski \cite{Dobzhansky-1973-ABT} captures the fact that the extraordinary diversity of life can only be understood by combining the mechanisms of genetic evolution with historical environmental threads. Consider now the common wisdom that, as a result of accumulated technological and financial innovations, societal and economic networks have never been more complex and that this complexity has reached unmanageable levels within the current understanding and methodologies \cite{Haldane-May-2011-Nature,Johnson-2011-Nature,Lux-2011-Nature}. Moreover, this complexity is often accused to be at the core origin of the financial crisis that started in 2007, of the ensuing so-called Great Recession and of the continuing woes of major economies worldwide. In the spirit of Dobzhanski's statement, we here propose to investigate the concept that nothing in the complexity of financial markets make sense except in the light of the evolution of investors' strategies and of their mutual feedback loops. Rather than fixating on so-called stylized facts \cite{Cont-2001-QF}, we propose to study the combined evolution of financial patterns  with the ecology of investors feeding on them and creating them. This is analogous to the importance of understanding the evolution of the fabric of social networks to make sense of the dynamics of human societies, the growth and organization of fault networks to account for the spatio-temporal organization of earthquakes, the structure of the brain and its plasticity to describe neural excitations and make progress on treating epileptic seizures, and so on. Similarly, the occurrence and severity of the financial crisis is best understood from the perspective of the accumulation of at least five bubbles over the last twenty years \cite{Sornette-Woodard-2009-XXX} associated with a climate of complacency everywhere and the illusion of the ``great moderation'' \cite{Bernanke-2004}.

The view that financial markets can be better understood as adaptive ecologies of co-evolving investors is not new. It has been explored in agent-based models \cite{LeBaron-Arthur-Palmer-1999-JEDC,Hommes-2001-QF,Hommes-2002-PNAS,Farmer-2002-ICC,Ehrentreich-2006-JEBO,Hommes-Wagener-2009} and articulated in the so-called ``adaptive markets hypothesis'' \cite{Lo-2004-JPM}. Here, our contribution is to provide novel empirical evidence based on the analysis of a unique dataset.

The logic of our approach is based on the following points.
\begin{enumerate}
\item Several important studies have shown that
efficient allocation can result from the aggregation of decisions made by irrational or zero-intelligent agents
under constraints  \cite{Gode-Sunder-1993-JPE,Othman-2008,Farmer-Patelli-Zovko-2005-PNAS}.

\item Many studies have repeatedly documented that
most investors underperform the global market as well as simple
buy-and-hold strategies \cite{Malkiel-2011,Barras-Scaillet-Wermers-2010-JF,Fama-French-2010-JF},
with only very few exceptions \cite{Kosowski-Timmermann-Wermers-White-2006-JF}.

\item  The structure of markets results from the aggregate impact of investors.

\item Here, we show that random strategies are as a rule significantly better than even the best investors.

\item We characterize the statistical properties of trading frequency and holding periods of investors
and quantify their impact on performance.
\end{enumerate}
Of course, point 4 is self-fulfilling in the sense that a random strategy is applied
to a system made by supposedly optimizing human beings who shape the market. If most strategies were random, the conclusions
are likely to be quite different.
Points 3-5  together imply that there are untapped investment and arbitrage opportunities,
that very few investors actually profit from.
This is very surprising given the ease with which
zero-intelligence traders over-perform. The quantitative characterization of the
performances of investors as a function of a few key observable characteristics that is presented below
can be used by future prospective investors to improve their strategies.
Of course, as more investors become wiser, the market characteristics will evolve
in a way similar to the first-entry games in which random strategies are found
ultimately to dominate \cite{Satinover-Sornette-2007-EPJB}.
Our main point is that the characterization
of financial markets requires understanding the ecology of strategies and their characteristics
and how they interact together to shape the very patterns they exploit.
Arguably, this will provide for really more efficient and robust financial markets,
designed to avoid future systemic crises.

In this work, we perform a statistical analysis of the performance of
all the investors trading 32 A-share stocks and 11 B-share stocks on the
Shenzhen Stock Exchange of China in 2003. This market
offers a unique opportunity
to compare (i) a closed national market (A-shares) with an international market (B-shares),
(ii) individuals and institutions and (iii) real investors to random strategies with respect
to timing that share otherwise all other characteristics.
 The analysis is conducted separately
for A-shares and B-shares. The database contains the information of each
order including (i) the masked ID of the trader, (ii) whether he is an
individual or institution, (iii) the direction, (iv) the price, (v) the size of the order,
and (vi) the time stamps accurate to 0.01 second
\cite{Gu-Chen-Zhou-2007-EPJB}. The evolution of the A-share index and
the B-share index is shown in Fig. S1. Interestingly,
the indices are found to be outperformed by the $1/N$ portfolio strategy
\cite{DeMiguel-Garlappi-Uppal-2009-RFS}. We find that the net return of
A-share individual investors is negative and independent of the trading
frequency, while that of A-share institutional investors, B-share
individual investors and B-share institutional investors decreases with
increasing trading frequency. In addition, the net return decreases for
winners and increases for losers when the trading frequency increases.
We also find that random trading performs better for all individuals and
institutions and for all winners. We show that the
performance of investors exhibit non-trivial power law dependence
as a function of trading frequency and holding periods.


\subsection*{The invisible hand with zero-intelligence agents}

Since Adam Smith's famous ``invisible hand'' description of
economic and financial markets as self-regulating systems, economics
has been dominated by the paradigm of rational utility
maximizing agents. Given restrictive conditions, the agents' collective actions
are found in theory to  lead to stable general equilibrium points that are characterized by
optimal allocation of resources. However, starting with H. Simon, and
expanding with the work of  D. Kahneman and A. Tversky  as well as many other scientists,
the severe limitations of human cognition and the many biases in real people's decisions have been pointed out.
These limitations and biases a priori cast doubts on the relevance of rational utility theory.
In reply, many studies have shown that irrational households can lead
in aggregate to rational markets.  In particular,
Gode and Sunder \cite{Gode-Sunder-1993-JPE} used ``zero-intelligence'' computer  agents
(who do not seek or maximize profits, do not observe, remember, or learn)
to simulate market transactions in a double auction. They found that a population
of such agents, subjected to budget constraint, produced results
that closely mirrored the allocation efficiency of a simultaneous
experimental human exchange. Studying prediction markets, Othman \cite{Othman-2008}
confirmed recently that prices that replicate the findings of empirical market studies
can emerge from a market populated by inhuman zero-intelligence agents with diffuse
beliefs. Farmer et al. \cite{Farmer-Patelli-Zovko-2005-PNAS} developed a model of zero-intelligent agents
that explains a large part of the cross-sectional properties of stocks
traded in continuous double auction markets. They suggest that
constraints imposed by market institutions may at times dominate strategic
agent behavior, so that random agents with constraints perform on the
whole as well as their more human siblings.

\subsection*{Underperformance and the illusion of control}

In both single-player and multiplayer Parrondo games, two or more losing games when alternated periodically or randomly yield a net winning outcome \cite{Parrondo-1996,Harmer-Abbott-1999-SS,Harmer-Abbott-1999-Nature,Harmer-Abbott-Taylor-Parrondo-2000-UPNF}. When an optimization rule is introduced, the Parrondo games produce degraded rather than enhanced returns \cite{Satinover-Sornette-2007-PA,Satinover-Sornette-2007-EPJB}. This ``illusion of control'' phenomenon is present in other agent-based models whose design is inspired by stock markets \cite{Satinover-Sornette-2009-EPJB}. The convention wisdom states that institutions markedly outperform individuals because they are more informed \cite{Barber-Odean-Zhu-2009-RFS}. However, the performance of both professionals and laymen is often documented to be worse than chance \cite{Torngren-Montgomery-2004-JBF,Barras-Scaillet-Wermers-2010-JF,Fama-French-2010-JF,Malkiel-2011}. Even worse, there is evidence showing that analysts' stock recommendation records are intentionally rewritten to a large extent \cite{Ljungqvist-Malloy-Marston-2009-JF}. Indeed, the performance of claimed successful strategies should be tested based on the method of random strategies, that are designed to remove survival and look-ahead biases \cite{Daniel-Sornette-Woehrmann-2009-JPM}.

The phenomenon of ``Illusion of control'' is one possible form of overconfidence. Overconfidence of stock market participants is expected to cause investors to trade more \cite{Odean-1998b-JF,Gervais-Odean-2001-RFS}, which has been confirmed at the market and individual equity level
\cite{Statman-Thorley-Vorkink-2006-RFS} and at the individual level \cite{Glaser-Weber-2007-GRIR,Glaser-Weber-2009-JFinM,Deaves-Luders-Luo-2009-RF}. In addition, there is evidence that the higher the frequency of trading, the poorer is the performance \cite{Odean-1999-AER,Barber-Odean-2000-JF,Barber-Odean-Zhu-2009-RFS,Barber-Lee-Liu-Odean-2009-RFS}.

\section*{Materials and Methods}
\label{S1:MaterialsMethods}

\subsection*{Data sets}
\label{S2:Data}

The Shenzhen Stock Exchange (SZSE) was established on December 1,
1990 and started its operations on July 3, 1991. It contains two
independent markets, A-share market and B-share market. The former
is composed of common stocks which are issued by mainland Chinese
companies. It is opened only to domestic investors, and traded in
CNY. The latter is also issued by mainland Chinese companies, while
it is traded in {\em{Hong Kong dollar}} (HKD). It was restricted to
foreign investors before February 19, 2001, and since then it has
been opened to Chinese investors as well. At the end of 2003, there
were 491 A-share stocks and 57 B-share stocks listed on the SZSE. In
the year 2003, the opening call auction is held between 9:15 am and
9:25 am, followed by the cooling periods from 9:25 am to 9:30 am,
and the continuous auction operating from 9:30 am to 11:30 am and
13:00 pm to 15:00 pm.

\begin{figure}[htb]
\centering
\includegraphics[width=8cm]{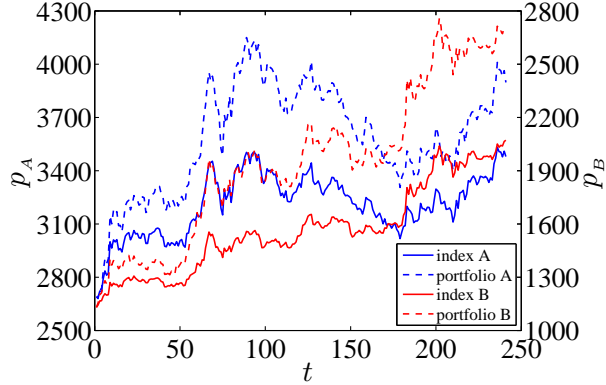}
\caption{\label{SFig:Index:evolution} Daily evolution of Shenzhen Component indexes for A-shares and B-shares in 2003 and the performance of their $1/N$ portfolios, respectively.}
\end{figure}

Our analysis is based on a database recording the order flows of 43
liquid stocks extracted from the A-share market and the B-share market on the SZSE in the
whole year of 2003 when the close call auction was adopted in the
opening procedure. The trading system did not show any information
about the order flows, and traders submitted orders only according
to the closing price of the last trading day. The database contains the
price, size and associated time of each submitted order recorded in
the opening call with the time stamps accurate to 0.01 second. Figure \ref{SFig:Index:evolution}
shows the evolution of the two indexes.

\subsection*{Method}
\label{S2:Method}

Assume that there are $I$ investors, each of them labeled $i$, and there are $S$ stocks in the database under investigation, each labeled $s$. For investor $i$ and stock $s$, we can construct a sequence of buy/sell activities, denoted $A(i,s)$:
\begin{equation}
 A(i,s) = \left[
 \begin{array}{ccccccccc}
 v_1, v_2,\cdots,v_j,\cdots, v_{J_s}\\
 p_1, p_2,\cdots,p_j,\cdots, p_{J_s}\\
 t_1, t_2,\cdots,t_j,\cdots, t_{J_s}
 \end{array}
 \right],
 \label{Eq:Ais}
\end{equation}
which means that investor $i$ {\bf{sells}} $v_j$ shares of stock $s$ with price $p_j$ at time $t_j$. If $v_j<0$, then ``sell $v_j$ shares'' means ``buys $-v_j$ shares''. Obviously, there are non-zero $v_j$ among all entries. We use the convention that
a positive sign corresponds to selling.

In order to construct $A(i,s)$, we need to reconstruct the order book. Assume that investor $i$ places a limit order of size $V$ at time $t$, which is executed later by $k$ other effective market orders with sizes $V_1, V_2, \cdots, V_k$ with prices $P_1, P_2, \cdots, P_k$ at times $T_1, T_2, \cdots, T_k$. Then, we record only one entry $(v_j,p_j,t_j)$, where
\begin{equation}
 p_j = \frac{1}{V}\sum_{m=1}^k V_mP_m,~~v_j=V,~~t_j = T_k.
 \label{Eq:pj:vj:tj}
\end{equation}
This is very important in the calculation of transaction costs, defined soon. More operations are needed for $A(i,s)$ in order to make sure that
\begin{equation}
  \sum_{j=1}^{J_s}v_j=0.
  \label{Eq:sum:vj}
\end{equation}
There are at least two cases to consider. First, if there are several sell transactions without any preceding buy transactions, these sells should not be included in $A(i,s)$. Second, at the end of year 2003, if investor $i$ holds some shares of stock $s$, we added a new entry by including a virtual transaction selling all his shares.

For the $j$-th transaction (or equivalently the $j$-th entry), the trading volume is $p_jv_j$. According to the Shenzhen Stock Exchange Trading Rules released in 2001, the transaction cost is determined as follows:
\begin{equation}
  c_j = \max\{|p_jv_j|(b_i+e+f), 5\} + |p_jv_j|d,
 \label{Eq:cj}
\end{equation}
The four terms in Eq.~(\ref{Eq:cj}) are the following: (i) Brokerage $b_i$, which should be less than 0.3\%; (ii) Exchange fee $e=0.01475\%$ for A-shares and $e_i=0.0301\%$ for B-shares for both buy and sell sides; (iii) Supervision fee $f=0.004\%$ for both buy and sell sides; and (iv) Stamp duty $d=0.1\%$ for sellers only. The sum of $b_i+e+f$ should be less than 0.3\% with a minimum of 5 CNY for A-shares and 5 HKD for B-shares for both buy and sell sides. Note that $b_i$ is $i$-specific and independent of stock $s$.

Therefore, the total invested capital (the money that investor $i$ spent to buy stock $s$) is $B_{i,s} = -\sum_{v_j<0} v_jp_j$ and the transaction cost of investor $i$ buying stock $s$ is $C_{i,s}^{b} = \sum_{v_j<0} c_j$. The total capital obtained by selling all the shares of stock $s$ is $S_{i,s} = \sum_{v_j>0} v_jp_j$ and the transaction cost of investor $i$ selling stock $s$ is $C_{i,s}^{s} = \sum_{v_j>0} c_j$.
The total transaction cost of investor $i$ in his investment of stock $s$ is
\begin{equation}
 C_{i,s} = C_{i,s}^{b}+C_{i,s}^{s},
 \label{Eq:Cis}
\end{equation}
The total earning is
\begin{equation}
 E_{i,s}=S_{i,s}-B_{i,s}-C_{i,s}+D_{i,s},
 \label{Eq:Eis}
\end{equation}
where $D_{i,s}$ is cash dividend received by agent $i$ from stock $s$ over the one
period. The portfolio return of investor $i$ can be calculated as follows
\begin{equation}
 R_i = \left.\sum_{s=1}^S E_{i,s}\right/\left(\sum_{s=1}^S B_{i,s}+\sum_{s=1}^S C_{i,s}^{b}\right).
 \label{Eq:Ri}
\end{equation}
The number of transactions (frequency of trading) is the sum of all $J$ values of investor $i$
\begin{equation}
 J_i = \sum_{s=1}^S J_{s}.
 \label{Eq:Ji}
\end{equation}

\section*{Results}
\label{S1:Results}

\subsection*{Basic statistics}

\begin{figure}[htb]
\centering
\includegraphics[width=8cm]{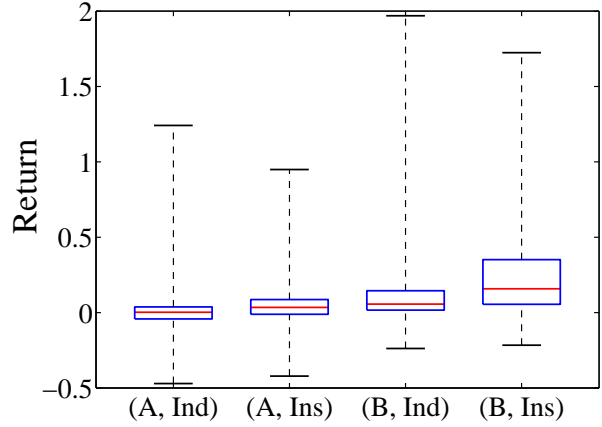}
\caption{Box plots of the returns in four classes: Individual A-share investors, institutional A-share investors, individual B-share investors, and institutional B-share investors. For each box plot, the minimum, lower quartile, median, upper quartile, and maximum of each class of returns are given. }
\label{Fig:Boxplot}
\end{figure}

In our database, there are 2,330,093 A-share investors with 2,315,664 individuals and 135,086 B-share investors with 88,779
distinct individuals. It is found that the proportion of institutional investors is much higher in the B-share market (34.28\%) than in the A-share market (0.62\%). For each investor $i$, we calculate his portfolio return $R_i$. For A-share investors, 51.95\% individuals and 68.50\% institutions are net winners above the zero benchmark. For B-share investors, 85.76\% individuals and 93.83\% institutions are winners
above the zero benchmark.
To have a full understanding of how many of the traders are really performing, the box plots of returns for each type of investors in each market are given in Fig.~\ref{Fig:Boxplot}. Some interesting observations are obtained.
\begin{itemize}
 \item The proportion of winning traders in the B-share market is much higher than in the A-share market.  This
 may be associated with the fact that the B-share index gained a much higher annual return than the A-share index (see Fig.~S1). Passive or even under-performing agents perform better
 on an absolue basis, the higher the upward trend of the underlying market.
 \item In both markets, the winning proportion of institutional traders is much higher than retail traders,
 which is consistent with recent results found for the Taiwan market \cite{Barber-Odean-Zhu-2009-RFS}.
 \item In each market, individual investors may gain higher returns than institutions or incur greater losses (see Fig.~\ref{Fig:Boxplot}).
\end{itemize}

\subsection*{Trading frequency and return}

\begin{figure}[htb]
\centering
\includegraphics[width=7cm]{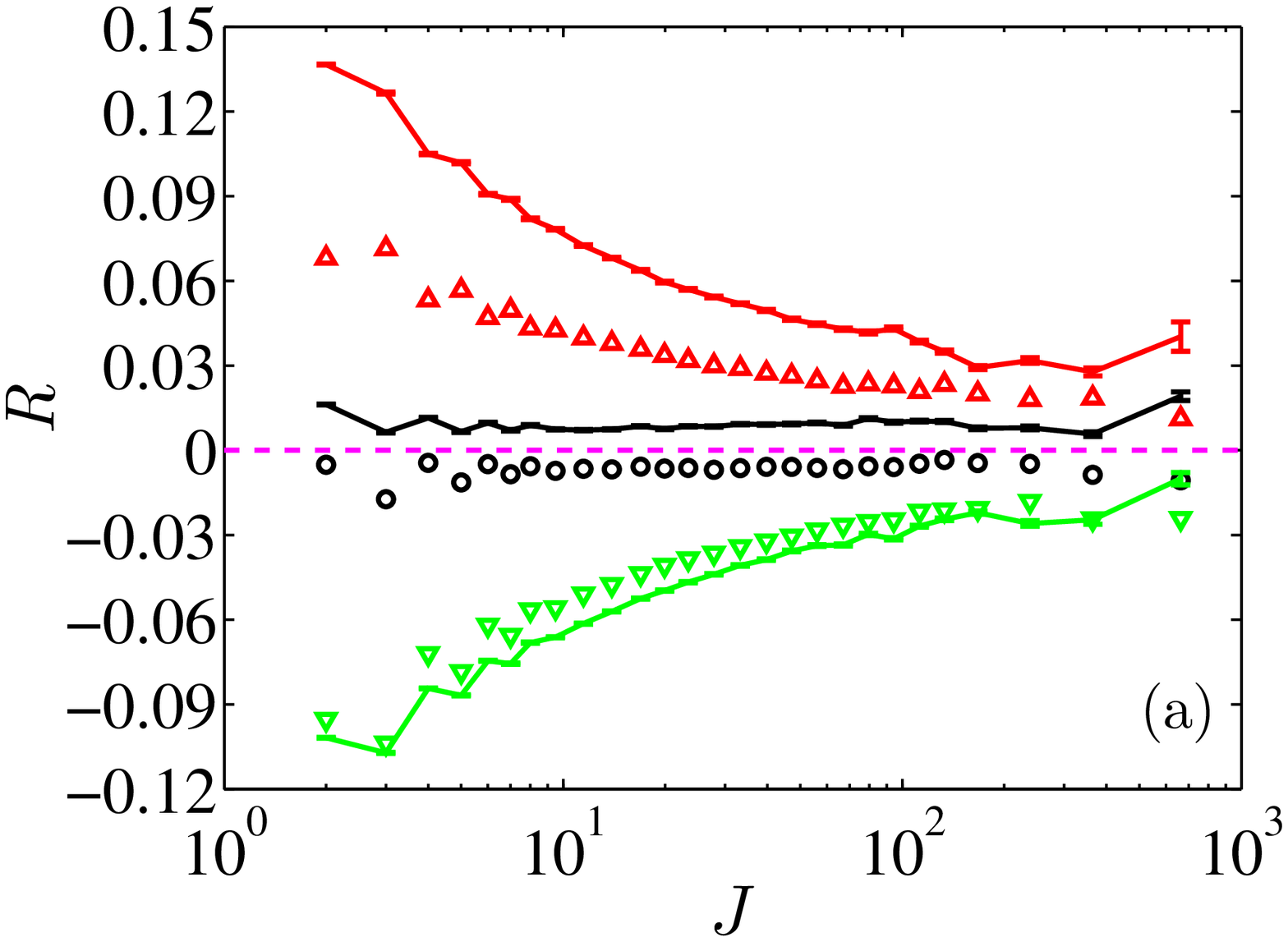}
\includegraphics[width=7cm]{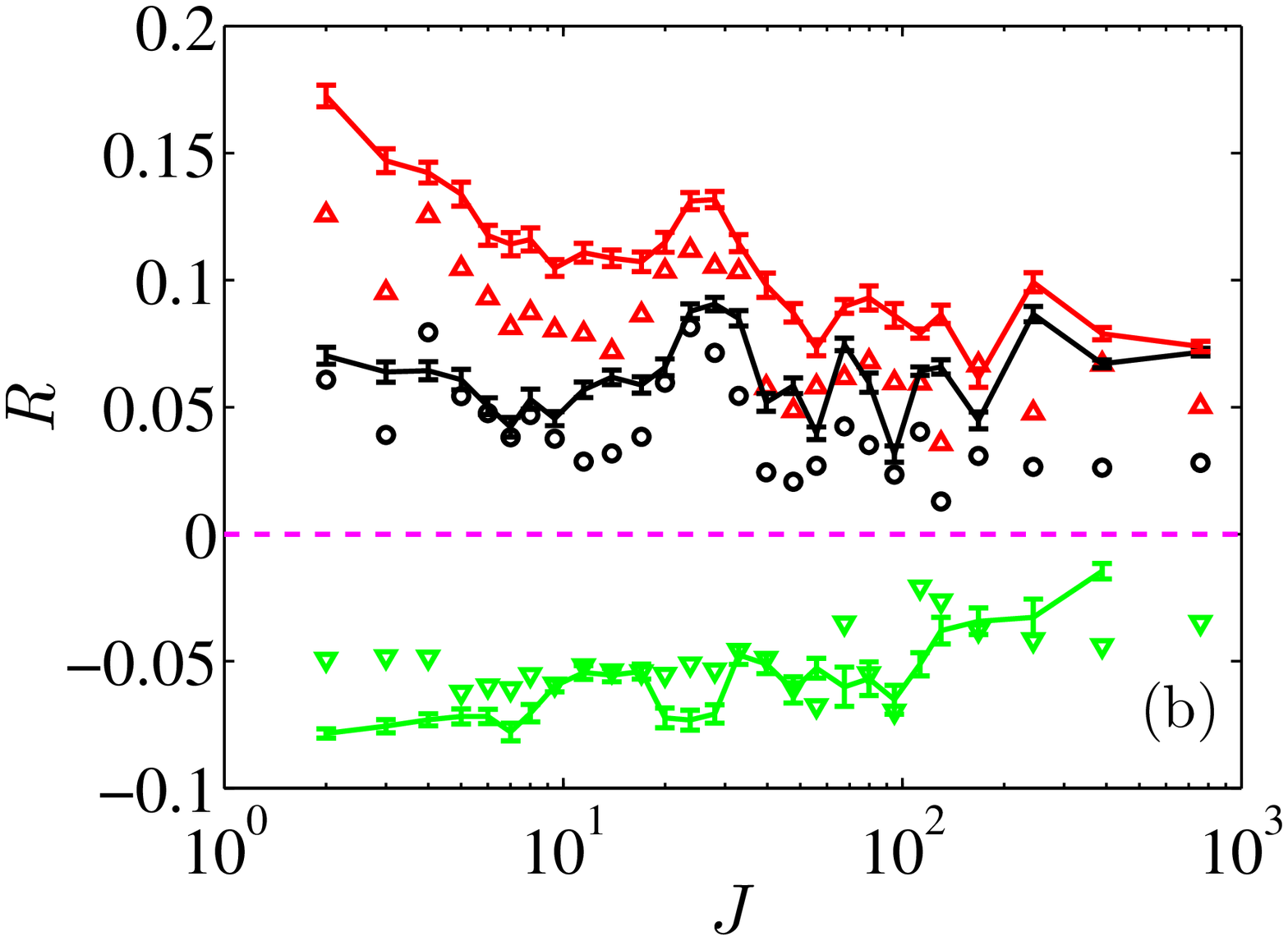}
\includegraphics[width=7cm]{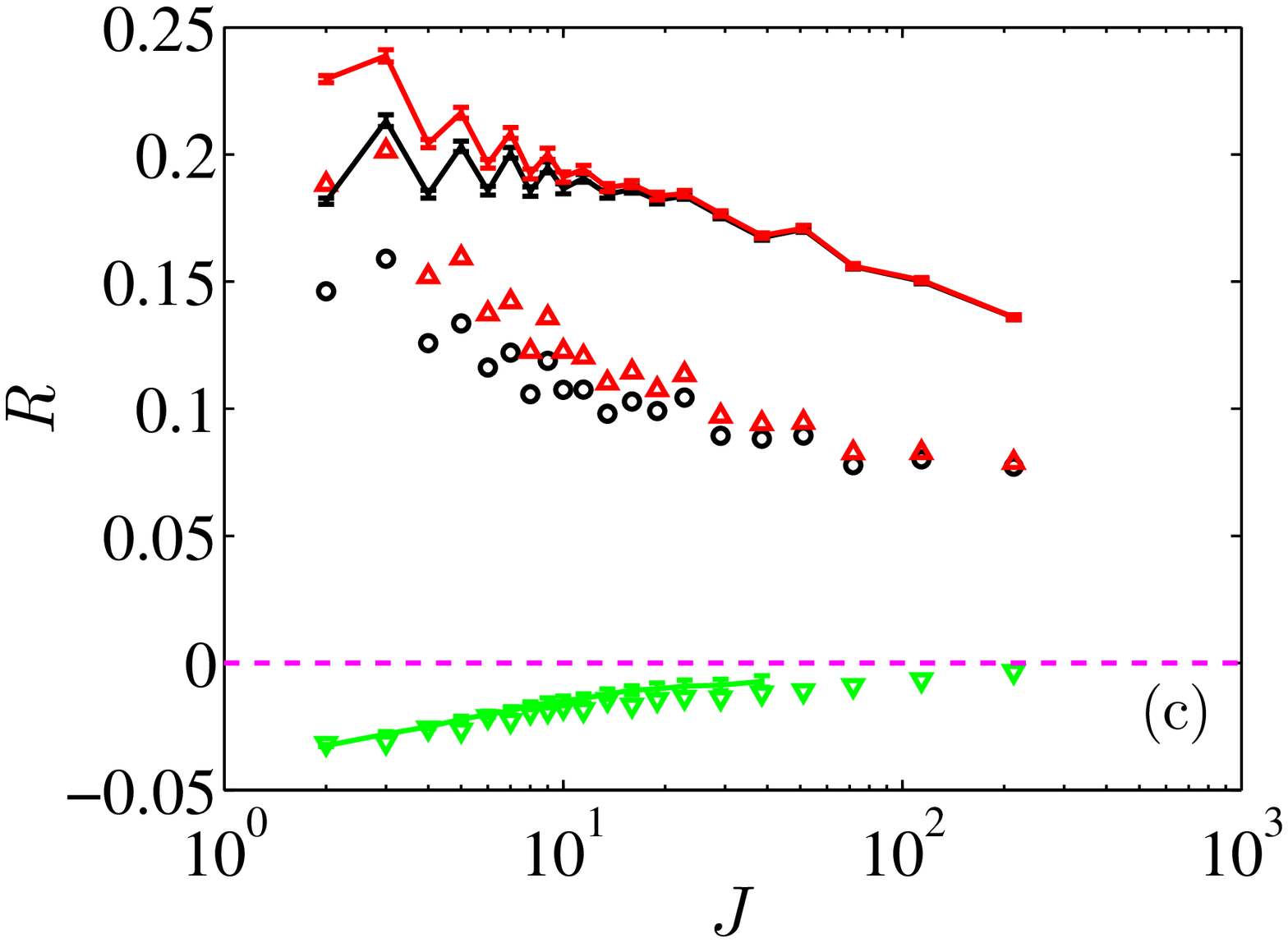}
\includegraphics[width=7cm]{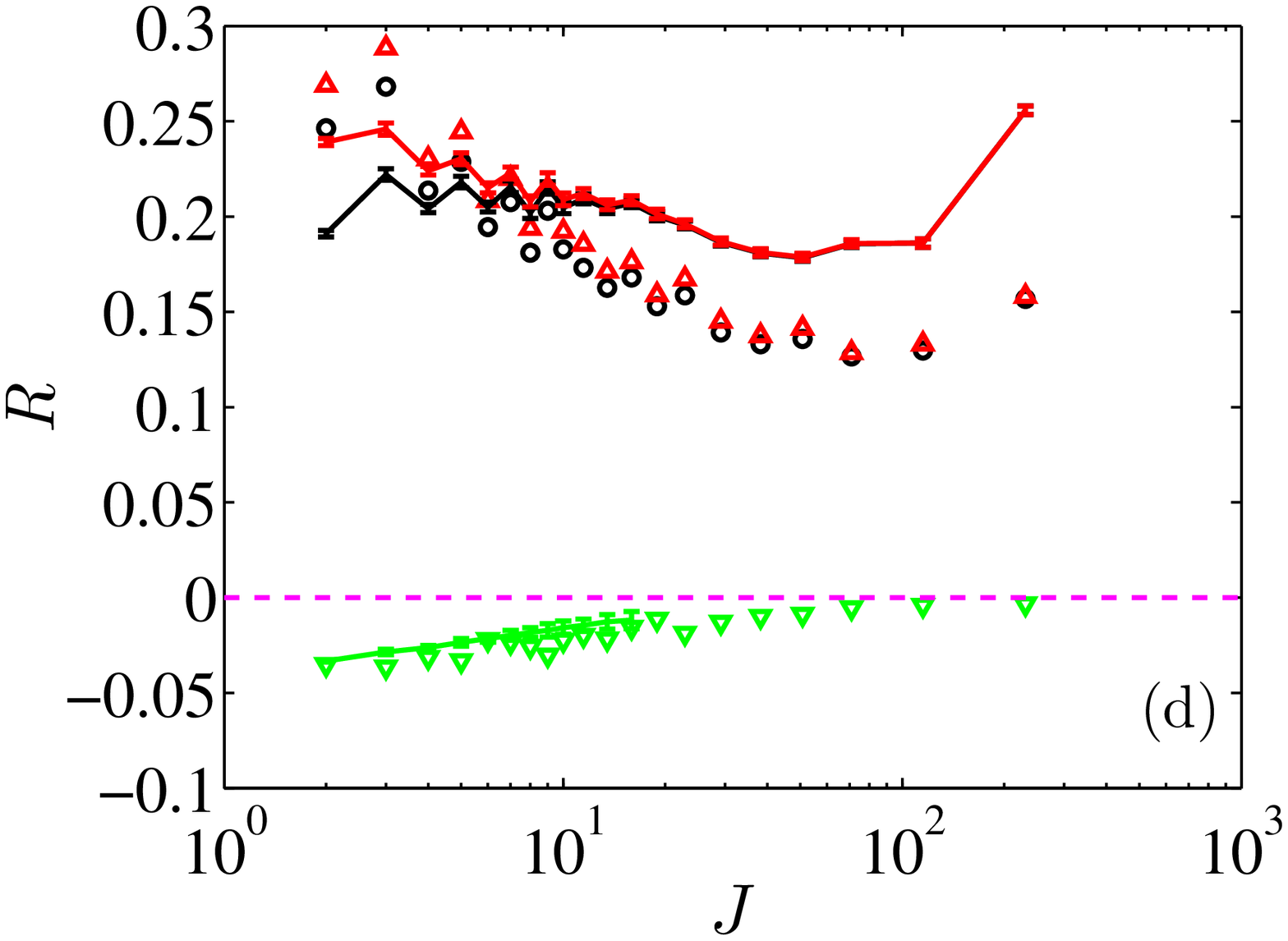}
  \caption{Performance comparison of strategic trading (real data) and random trading using the average values of return $R$ versus trading frequency $J$. We exclude the sell transactions without any preceding matching buys. The simulations for the random strategies are repeated for 2000 times. We show the results for individuals in A-share (a), institutions in A-share (b), individuals in B-share (c) and institutions in B-share (d), respectively. In each plot, the colorful symbols ({\color{black}{$\circ$}}, {\color{red}{$\vartriangle$}}, {\color{green}{$\triangledown$}}) correspond to strategic trading, the continuous lines correspond to random trading, and the dashed line indicates the base line of zero return ($R=0$).}
\label{Fig:Ave:RJ:net:all}
\end{figure}

Figure \ref{Fig:Ave:RJ:net:all} shows the dependence of the average returns $R$ as a function of the trading frequency $J$ for A-share individuals (a), A-share institutions (b), B-share individuals (c), and B-share institutions (d), respectively. One can observe that the return is statistically independent of the trading frequency for A-share individuals while, in the three other cases, $R$ decreases systematically with $J$, indicating that trading is hazardous to investors' wealth not only for individuals but also for institutions \cite{Barber-Odean-2000-JF}. Comparing the returns with the same trading frequency, institutions out-perform individuals.

We also investigate the dependence of the net return as a function of trading frequency for two categories of investors, the winners and the losers. Winners (respectively losers) are defined as those having a positive (respectively negative) return. The classification is thus performed on an absolute (and not relative) basis. Figure \ref{Fig:Ave:RJ:net:all} shows that return decreases with trading frequency for winners and increases for losers. The enhanced performance of losers by increasing trading frequency cannot be explained by a learned overconfidence bias as described in Refs.~\cite{Odean-1998b-JF,Gervais-Odean-2001-RFS}.

Figure \ref{Fig:Ave:RJ:net:all} also presents the average returns that random trading would yield in these markets. Random strategies are generated by considering each investor individually in turn, choosing random times for their trades while otherwise keeping fixed all other characteristics such as his number of transactions (trading frequency) and the trade sizes on each stock. Specifically, in Eq.~(\ref{Eq:Ais}), for a given investor $i$ and a stock $s$, the variables $J_s$ and $v_j,j=1,\cdots,J_s$ are unchanged, while the times $t_j,j=1,\cdots,J_s$ are replaced by a randomly chosen time sequence. As a result, the prices $p_j$ are also changed. This is done 2000 times for each investor,  generating overall a very large number synthetic outputs contributed over all the investors in our database. For a given frequency $J$, we sort again these many outputs into two classes: (i) the winners are the random strategies with a positive return; (ii) the losers are the random strategies with a negative return. We then compute separately the average returns (and their standard deviation) of the winning and of the losing random strategies, as well as the overall average return of these random strategies. Note that this construction of random strategies tests specifically the skills of investors with respect to timing, since all the other characteristics are kept otherwise identical \cite{Daniel-Sornette-Woehrmann-2009-JPM}. According to Fig.~2, the aggregate net return of random trading (black lines) is higher than that of real trading (black circles) with the same trading frequency in every case. More impressively, the aggregate net returns of A-share individuals are negative, while the net returns using random trading strategy are positive. Two closely related conclusions can be drawn: (i) real trading is not random but strategic; (ii) the performance of strategic trading is worse than random trading. For the winners, random trading also induces higher return than real strategic trading in all four cases. For losers in the A-share market, random trading performs slightly worse. For losers in the B-share market, the random trading and real trading yield almost identical net returns.

\subsection*{Holding time and return}

\begin{figure}[htb]
\centering
 \includegraphics[width=7cm]{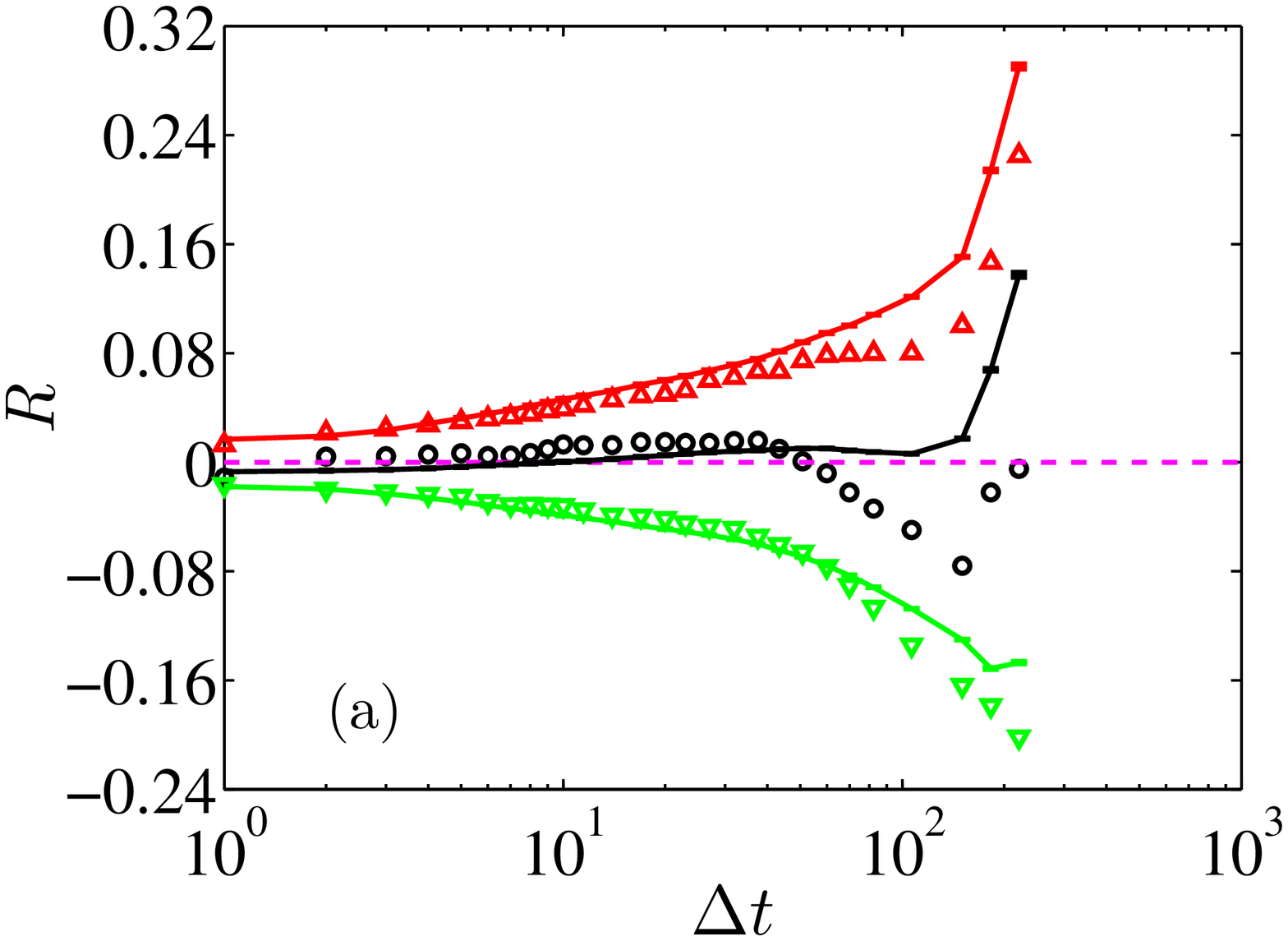}
 \includegraphics[width=7cm]{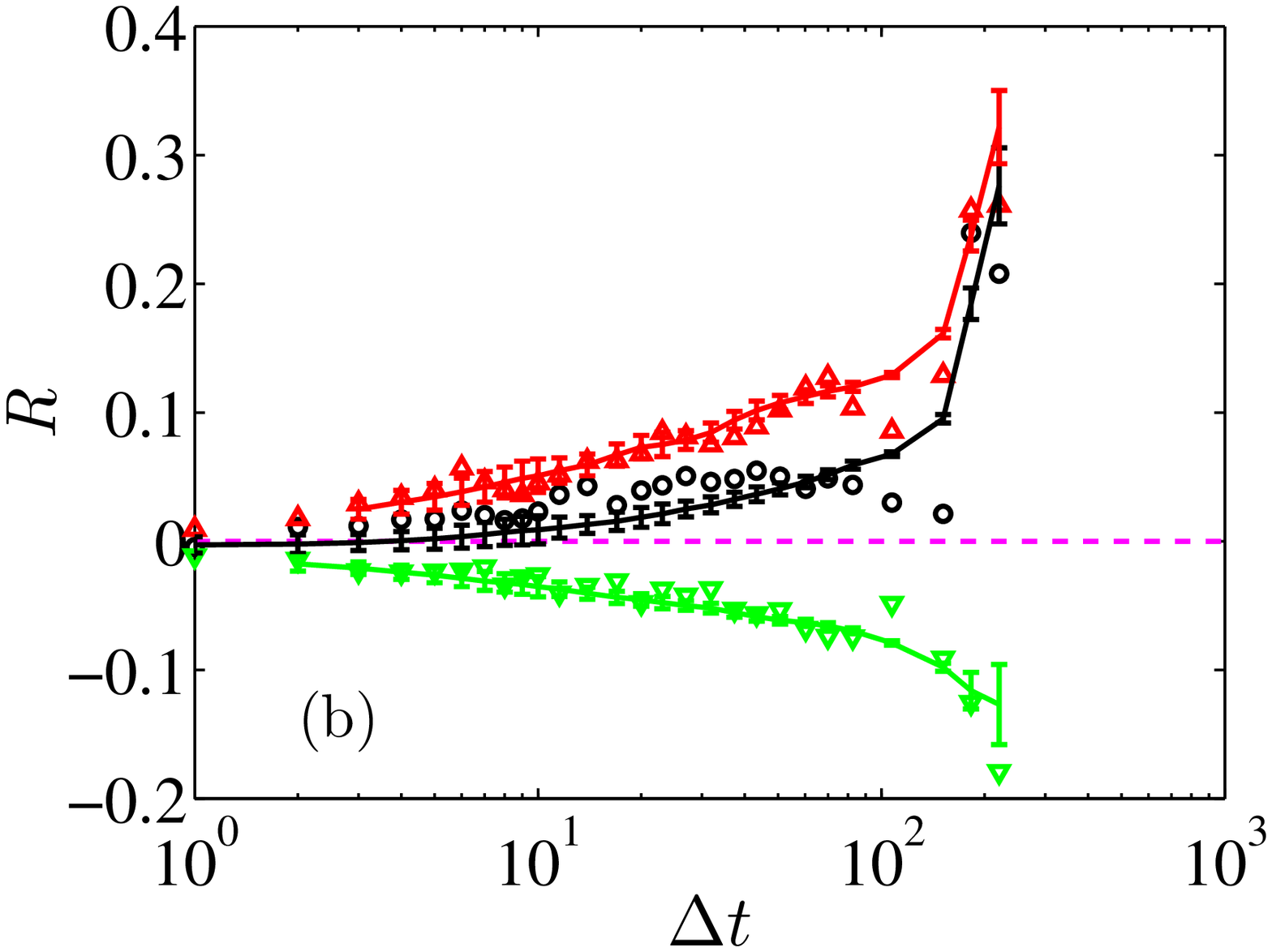}
 \includegraphics[width=7cm]{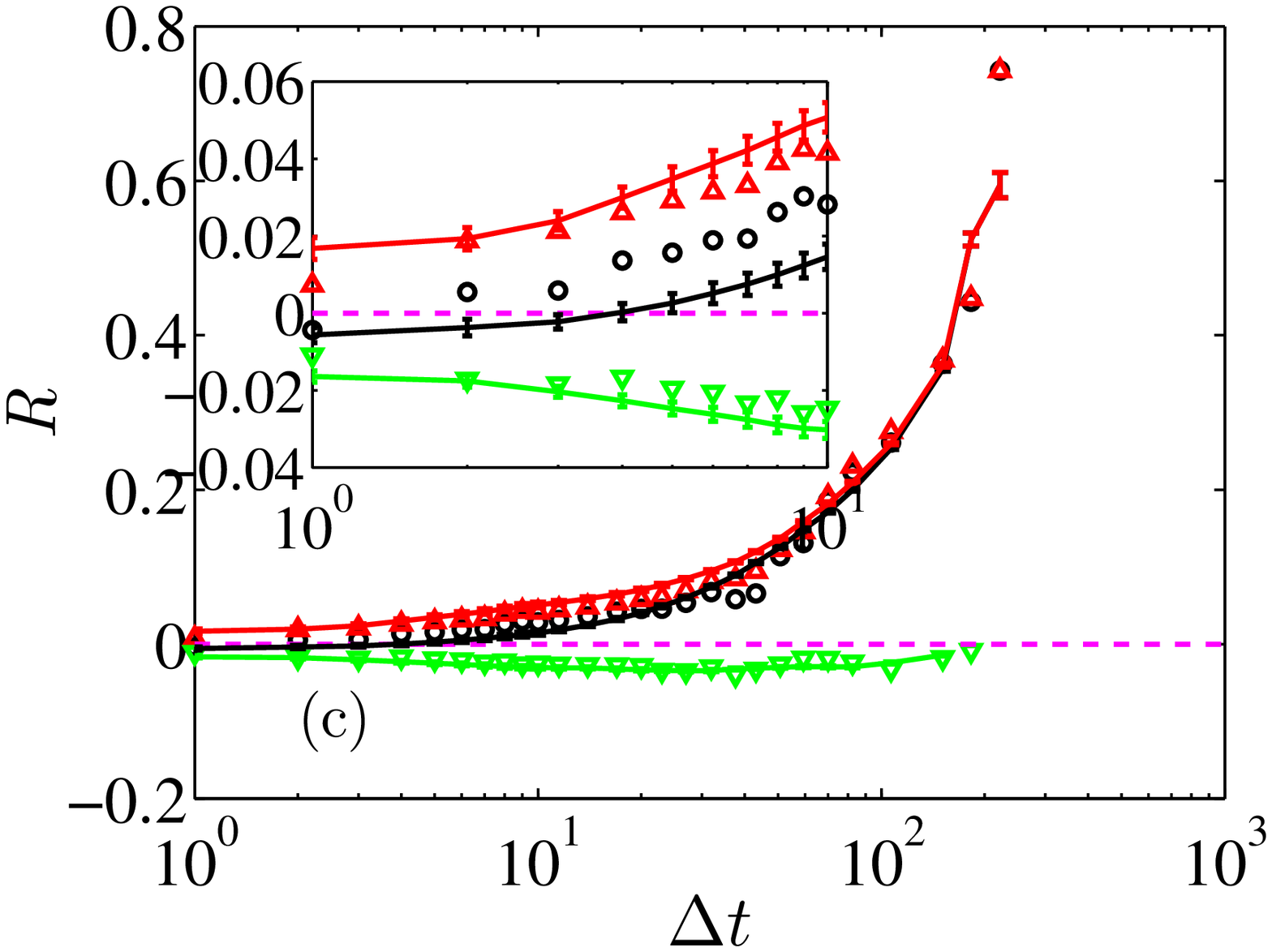}
 \includegraphics[width=7cm]{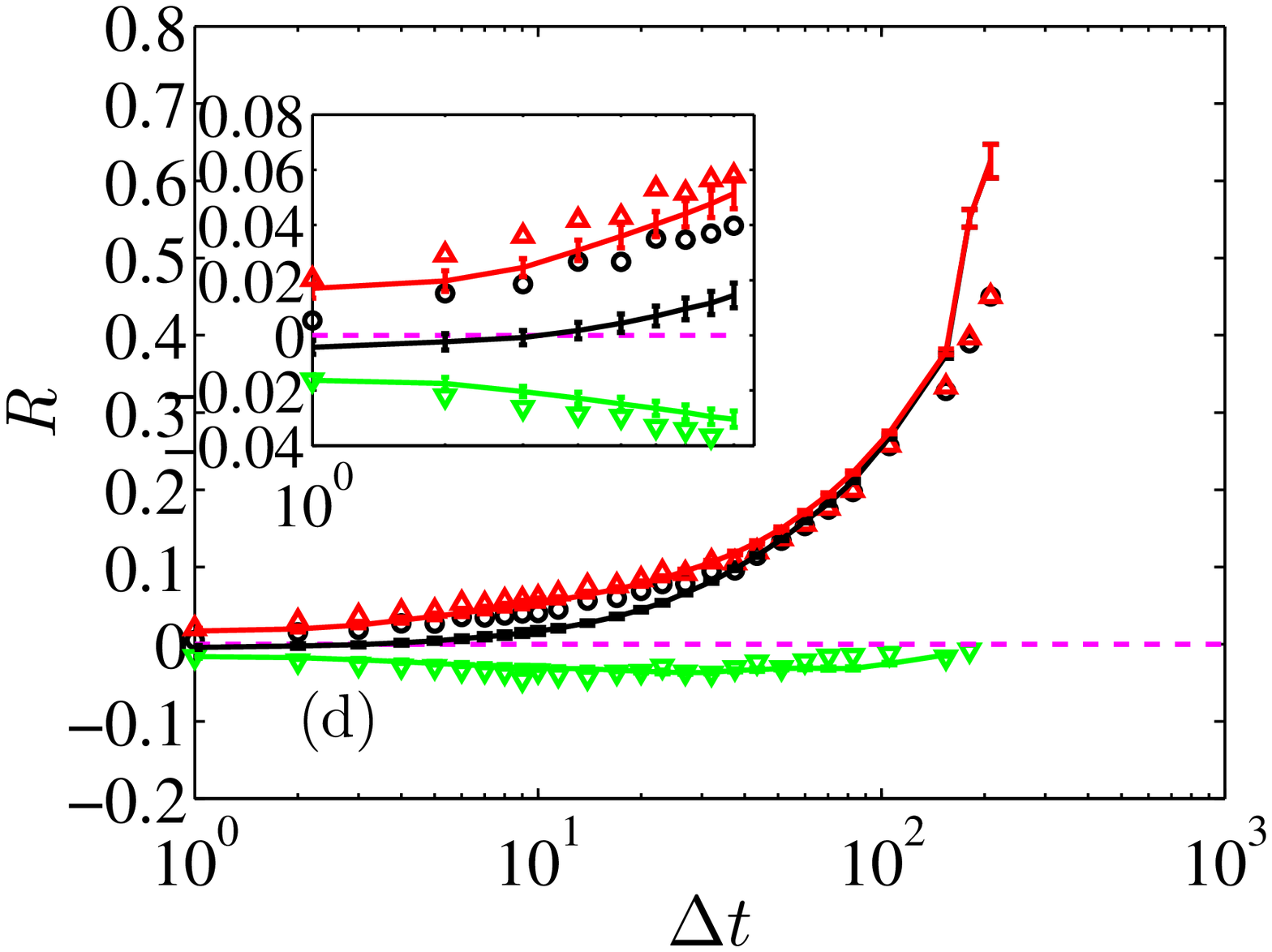}
 \caption{Average returns $R$ versus average holding time for individuals in A-share (a), institutions in A-share (b), individuals in B-share (c) and institutions in B-share (d), respectively. The symbols present the average values over all investors ($\circ$), as well as investors who earn positive return ({\color{red}{$\vartriangle$}}) and negative return ({\color{green}{$\triangledown$}}). The dashed line delineates
 the benchmark in absolute terms of zero return. The solid lines with error bars are the average simulation results of random trading.
 The insets in panels (c) and (d) are magnifications of the curves for small values of $\Delta t$.}
 \label{Fig:Ave:RT:net:all}
\end{figure}

Figure \ref{Fig:Ave:RT:net:all} shows the average return $R$ as a function of the average holding time $\Delta{t}$ for A-share individuals (a), A-share institutions (b), B-share individuals (c) and B-share institutions (d), respectively. This figure is different from Fig.~\ref{Fig:Ave:RJ:net:all} because holding time is not simply the inverse of trading frequency. Indeed, the total holding time $J\Delta{t}$ is not constant for different investors. In general, the aggregate net return increases with average holding time. This is consistent with the conventional wisdom that the buy-and-hold strategy outperforms most other strategies (see Fig.~S1 for the performance of the buy-and-hold strategy of the $1/N$ portfolio). For winners, the return is large when the holding time is long. For losers, $R$ decreases with respect to $\Delta{t}$ in the A-share market and varies slightly in the B-share market. The solid lines with error bars are the average simulation results of random trading. For large $\Delta{t}$ and B-shares, we still see that random trading performs better.

\subsection*{Quantitative relations linking return to trading frequency and holding period}

\begin{figure}[h!]
\centering
\includegraphics[width=5cm]{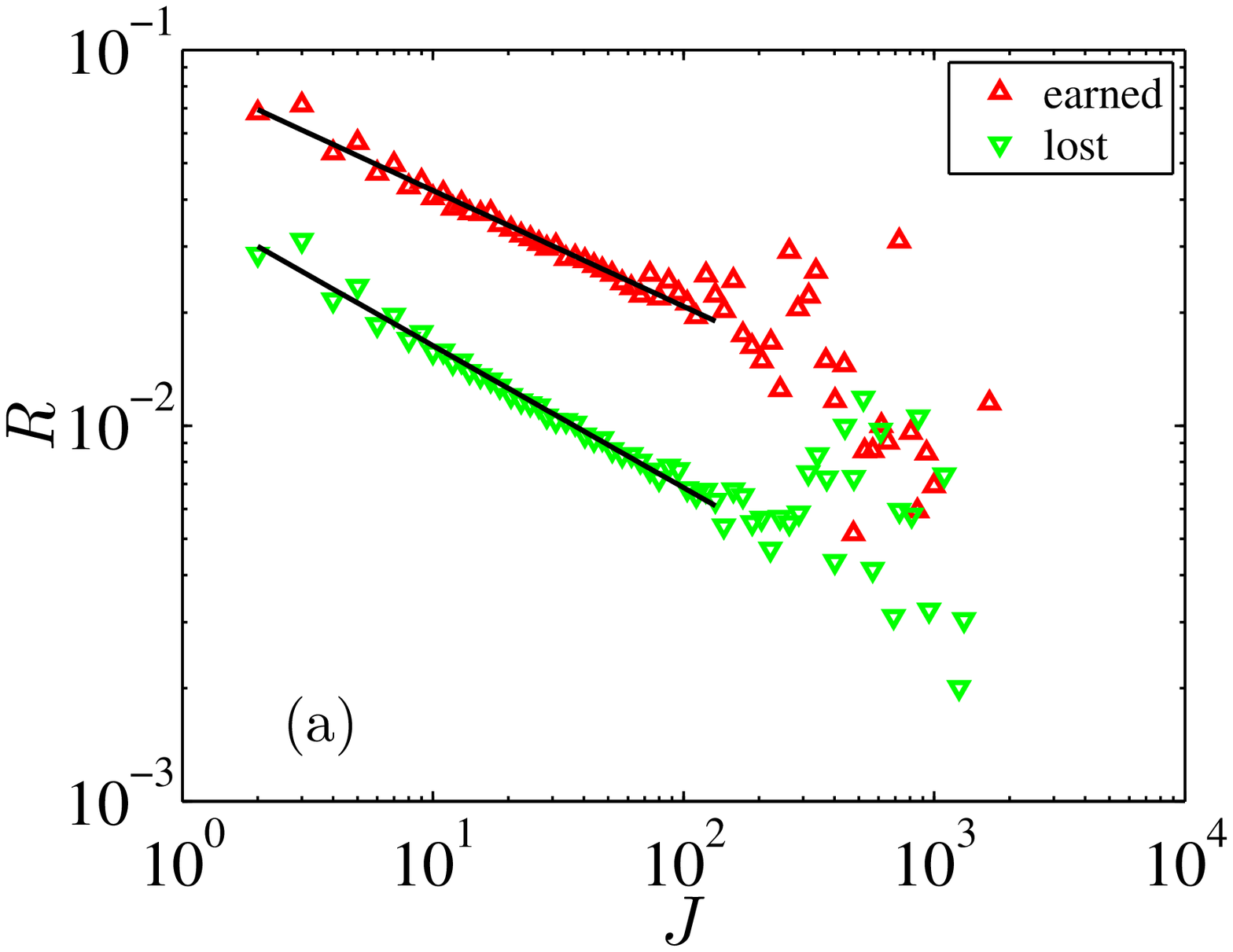}
\includegraphics[width=5cm]{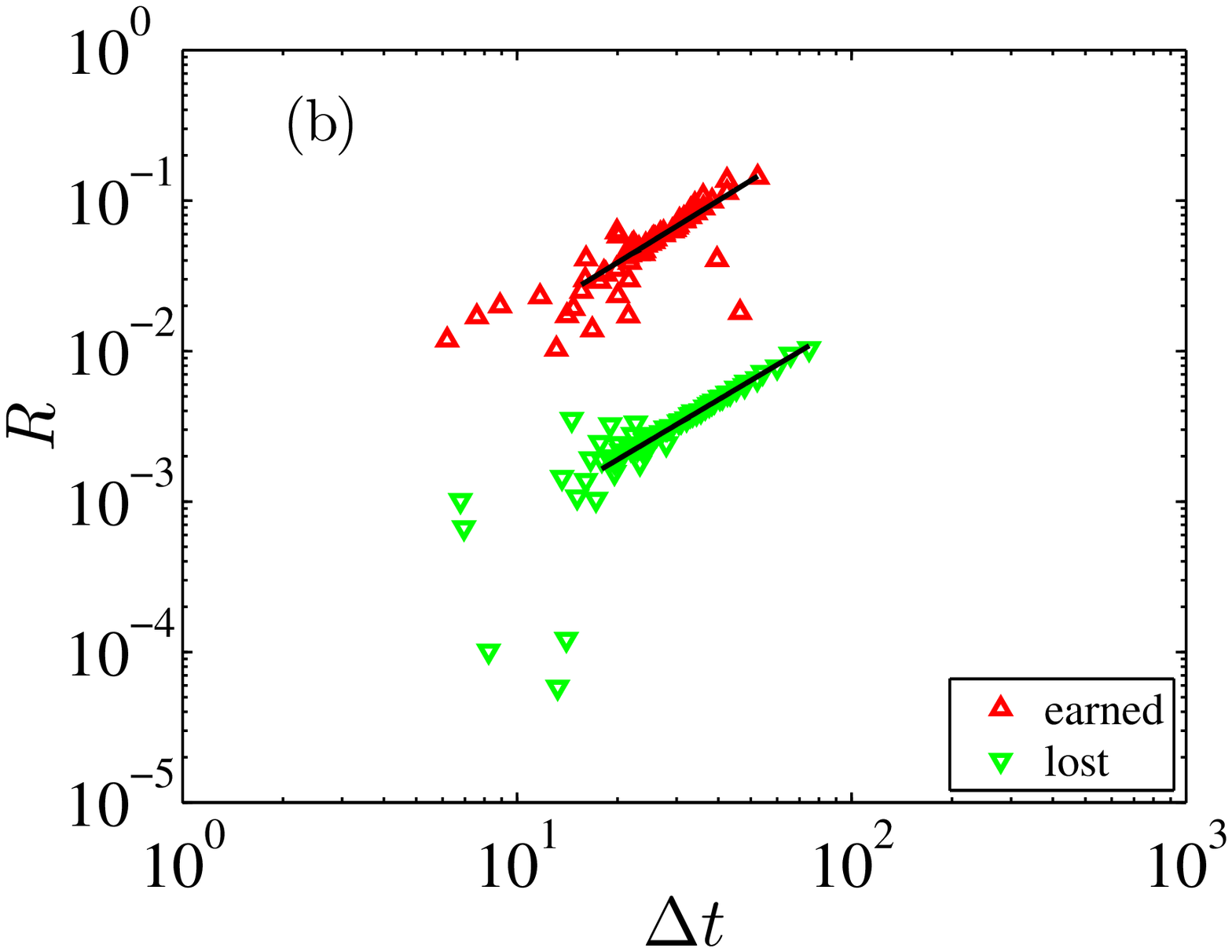}
\includegraphics[width=5cm]{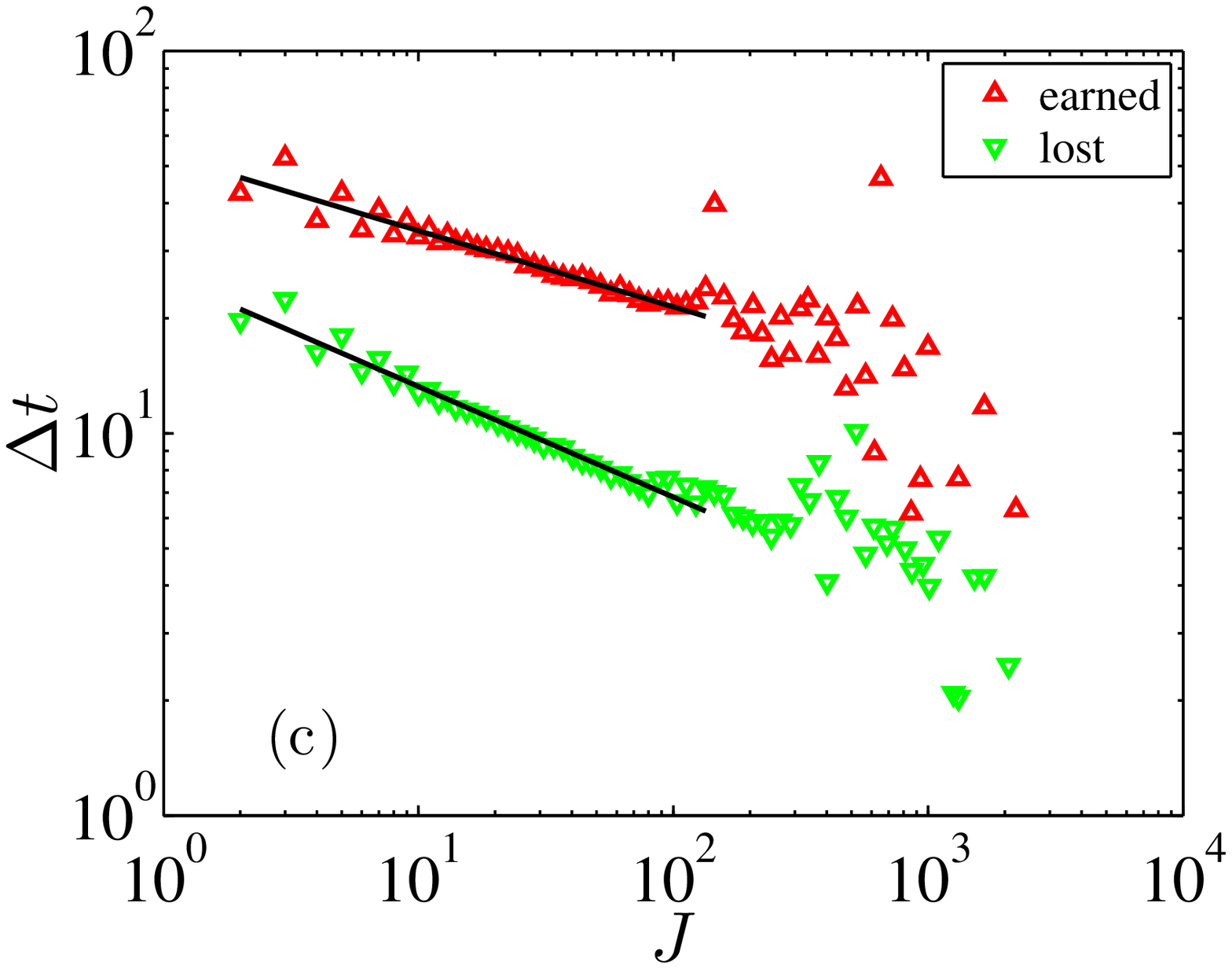}
\includegraphics[width=5cm]{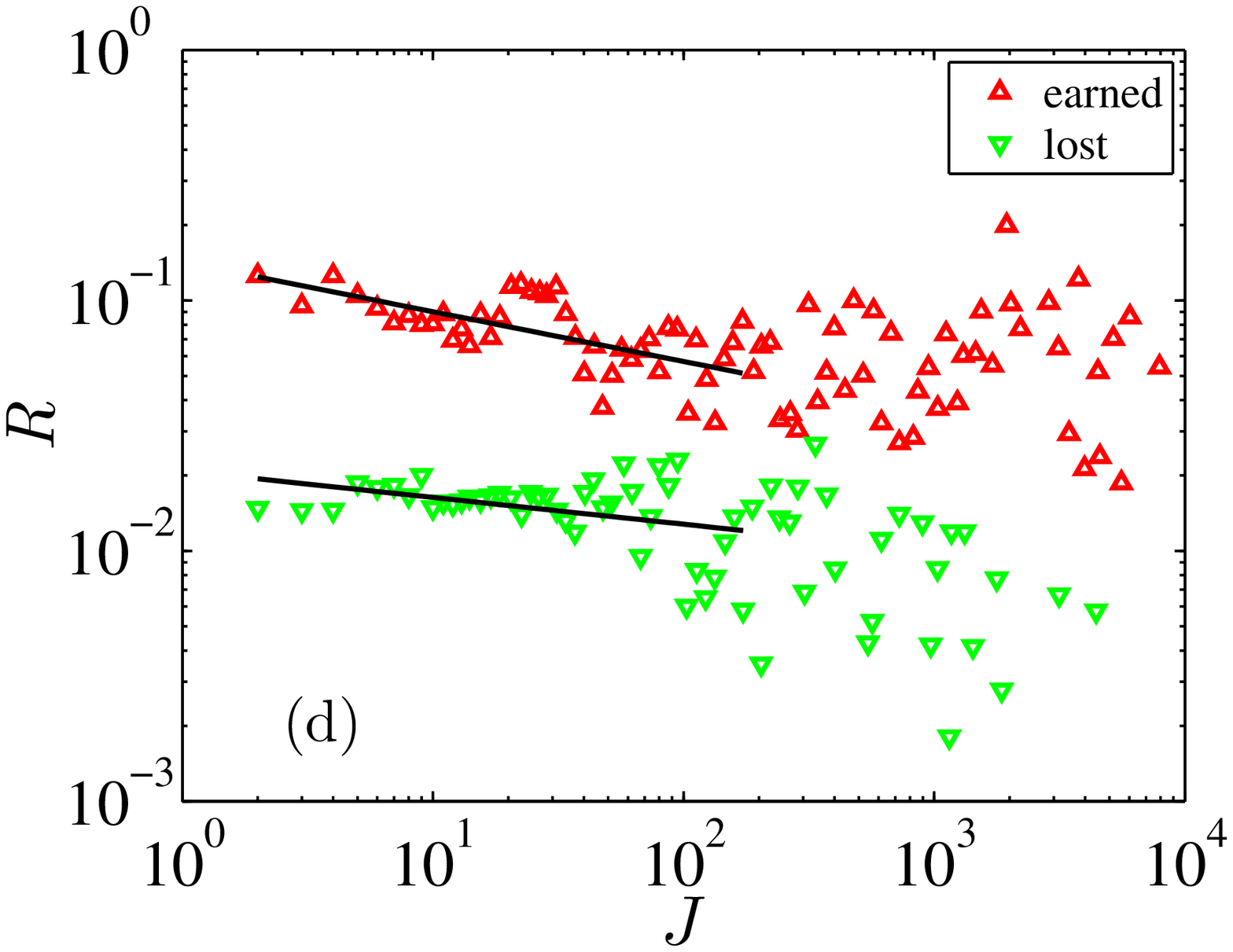}
\includegraphics[width=5cm]{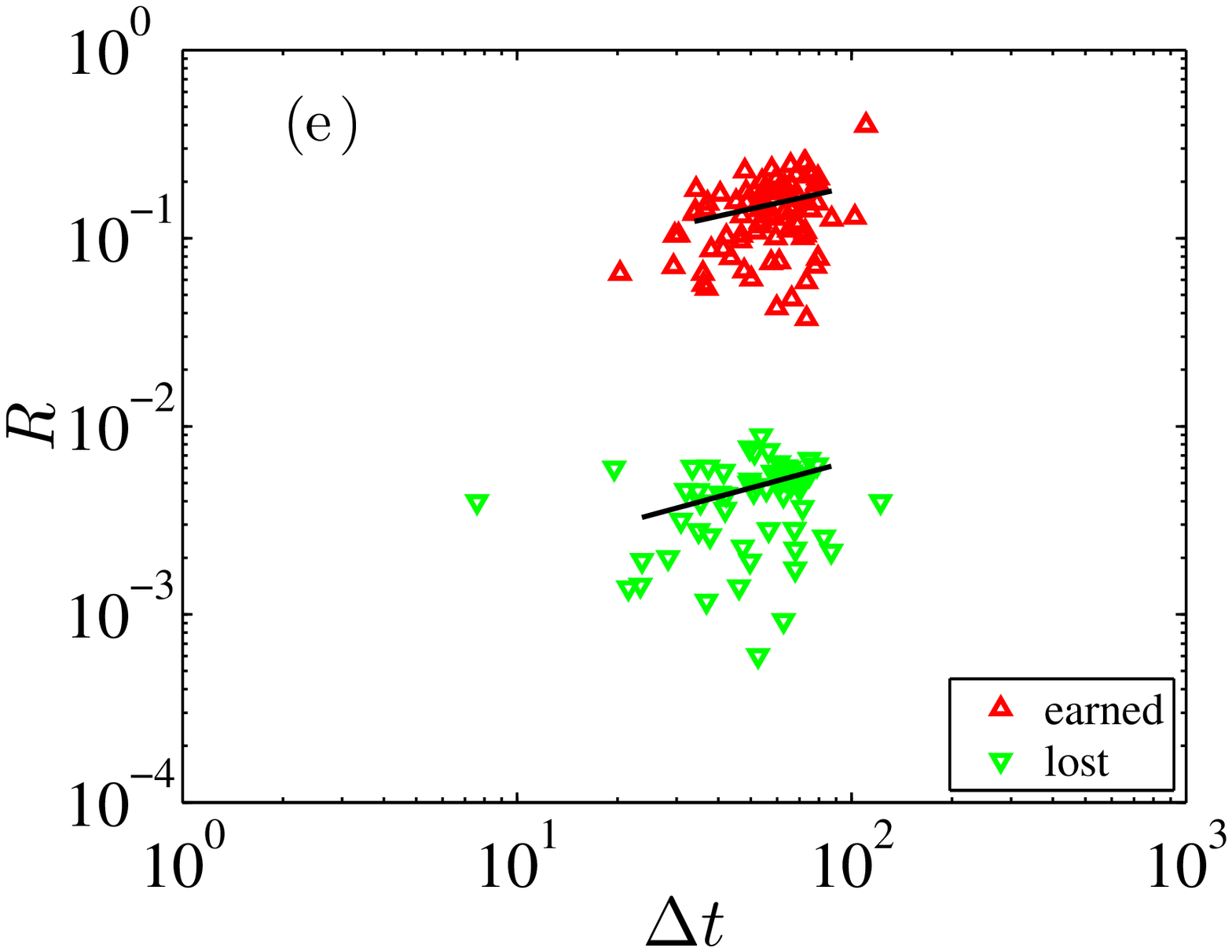}
\includegraphics[width=5cm]{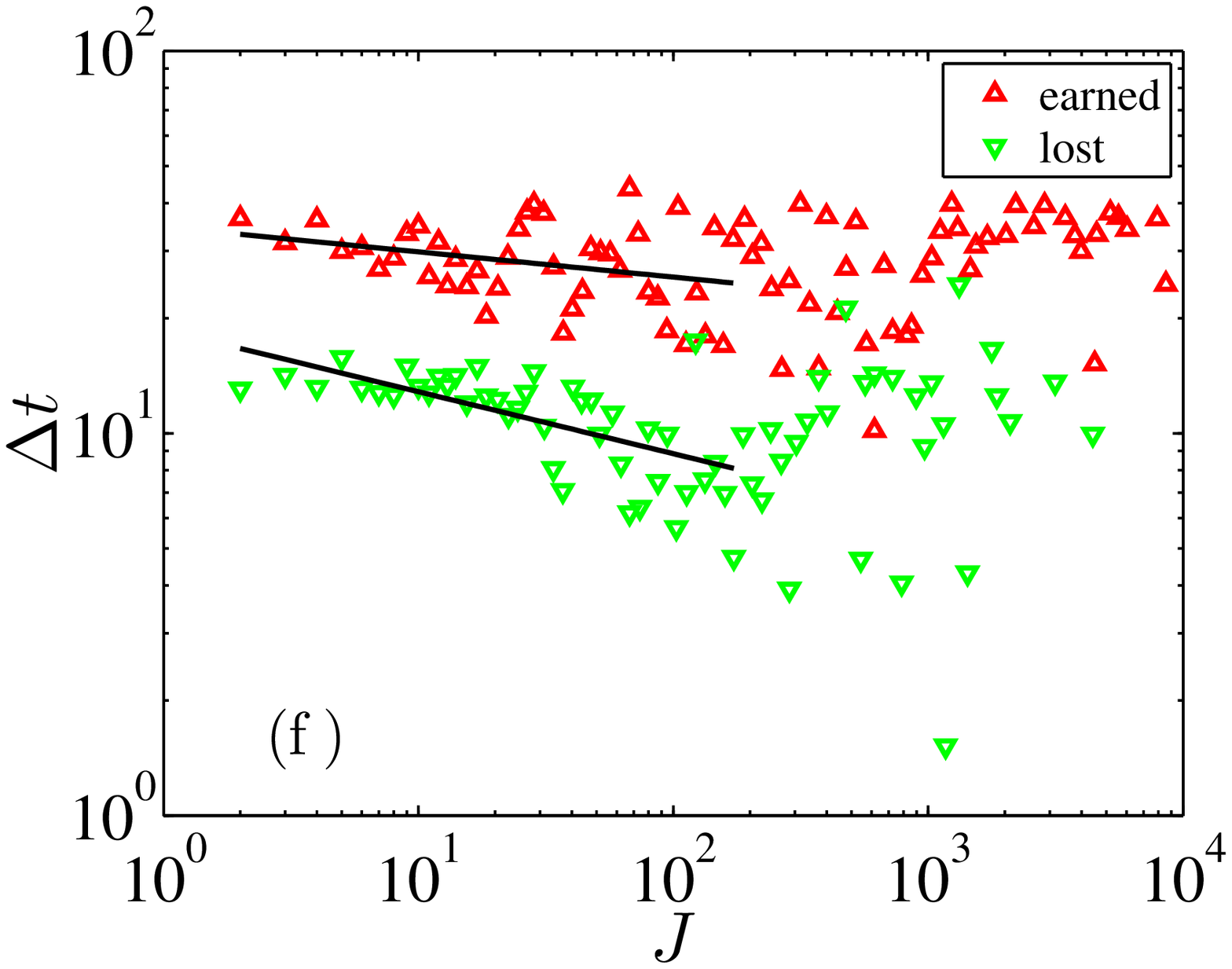}
\includegraphics[width=5cm]{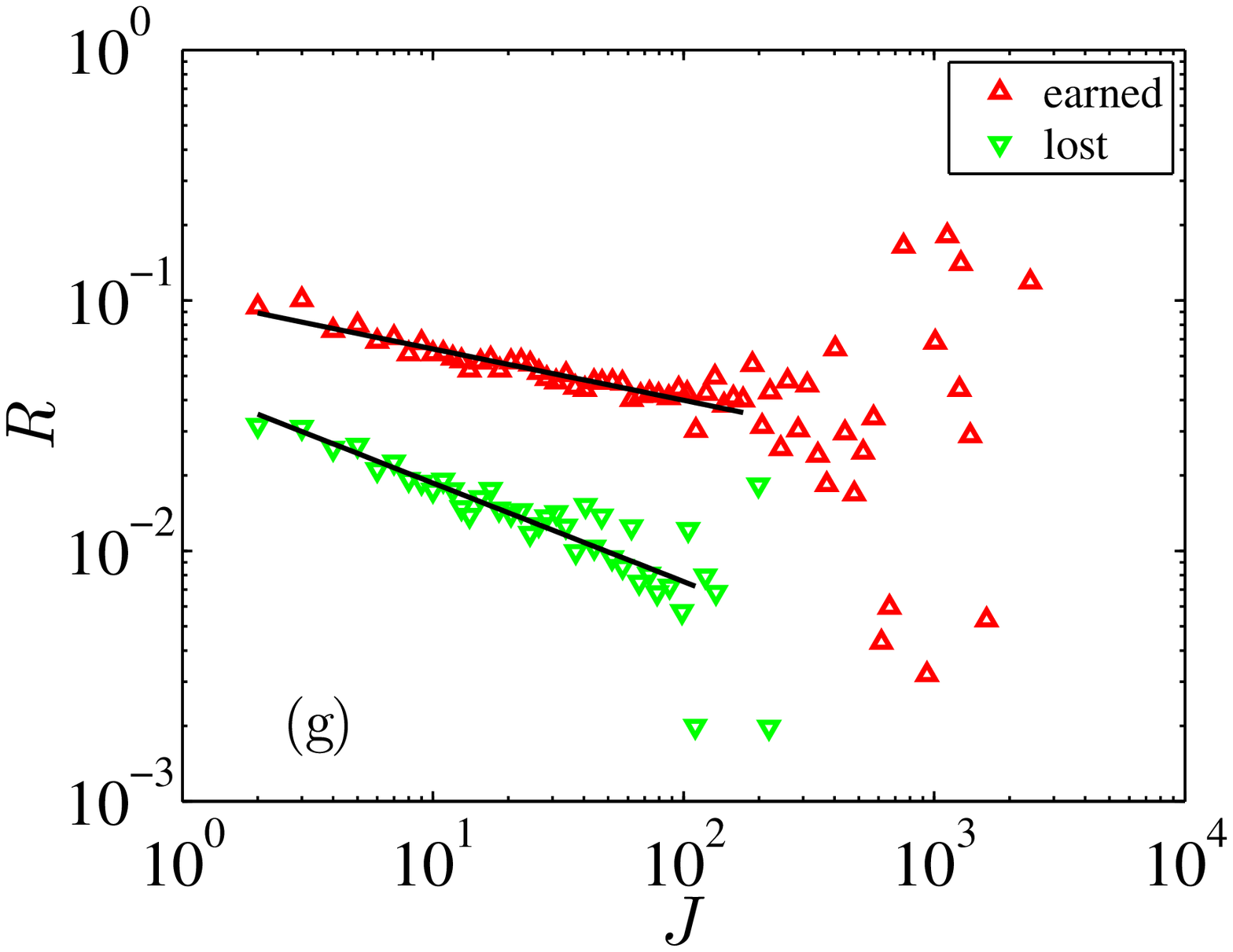}
\includegraphics[width=5cm]{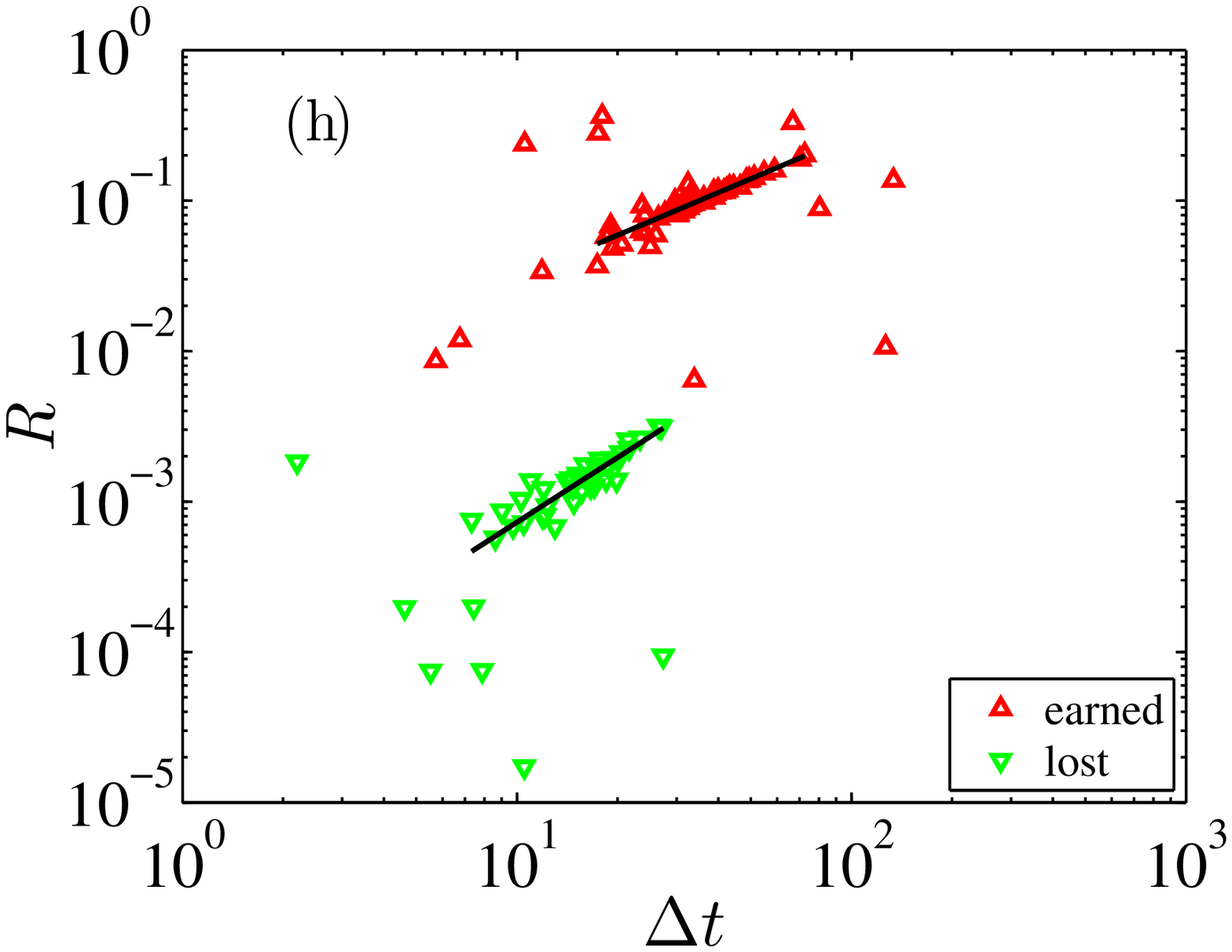}
\includegraphics[width=5cm]{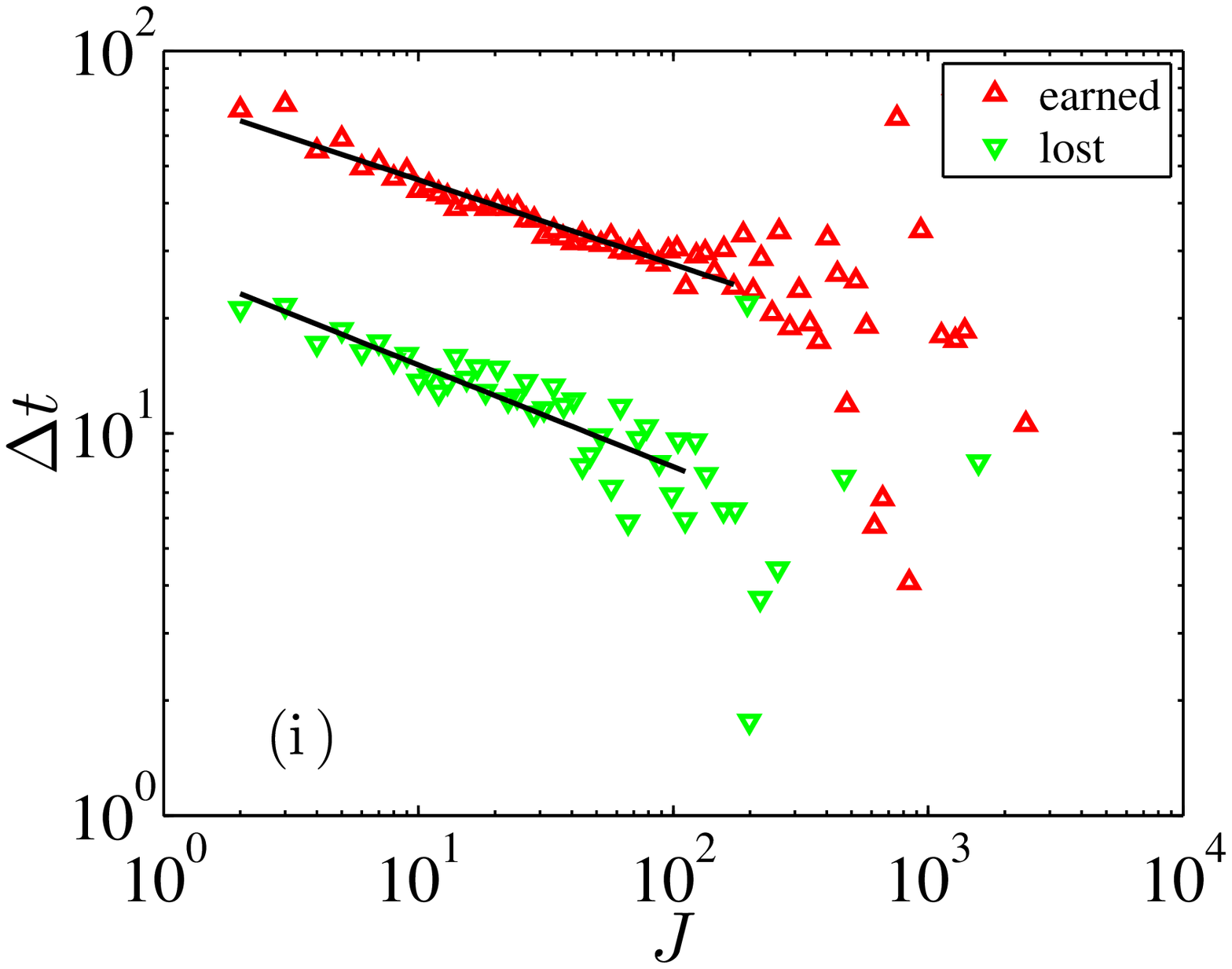}
\includegraphics[width=5cm]{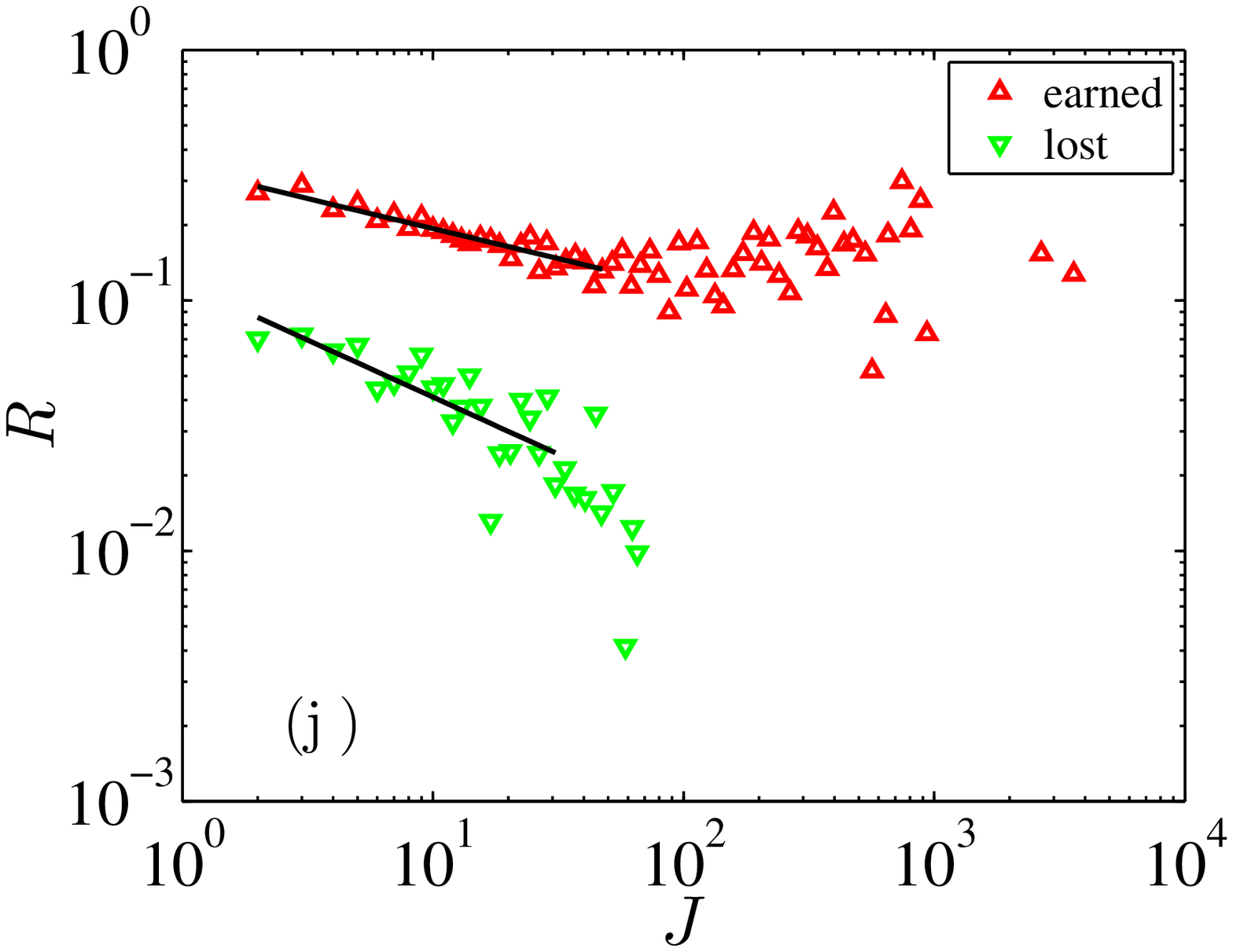}
\includegraphics[width=5cm]{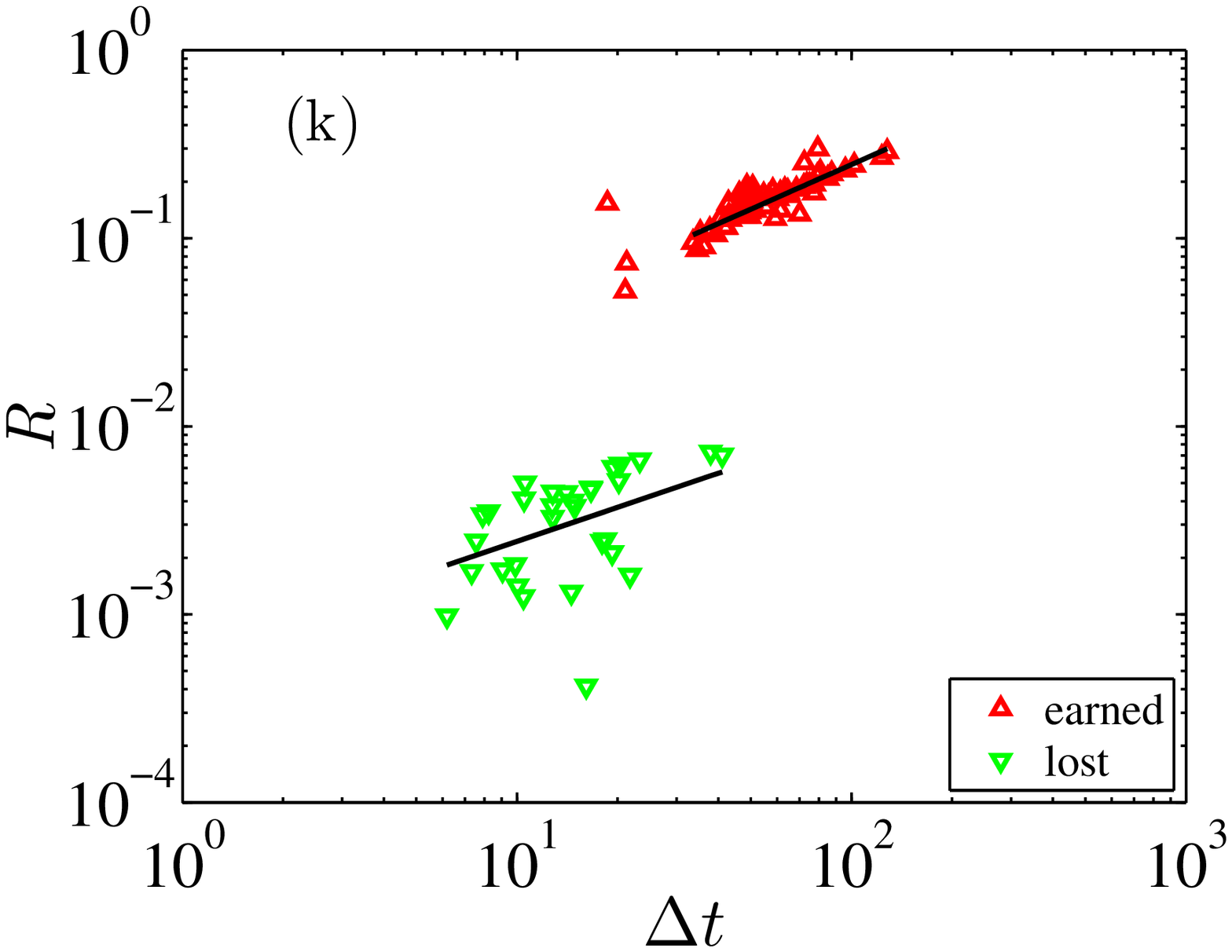}
\includegraphics[width=5cm]{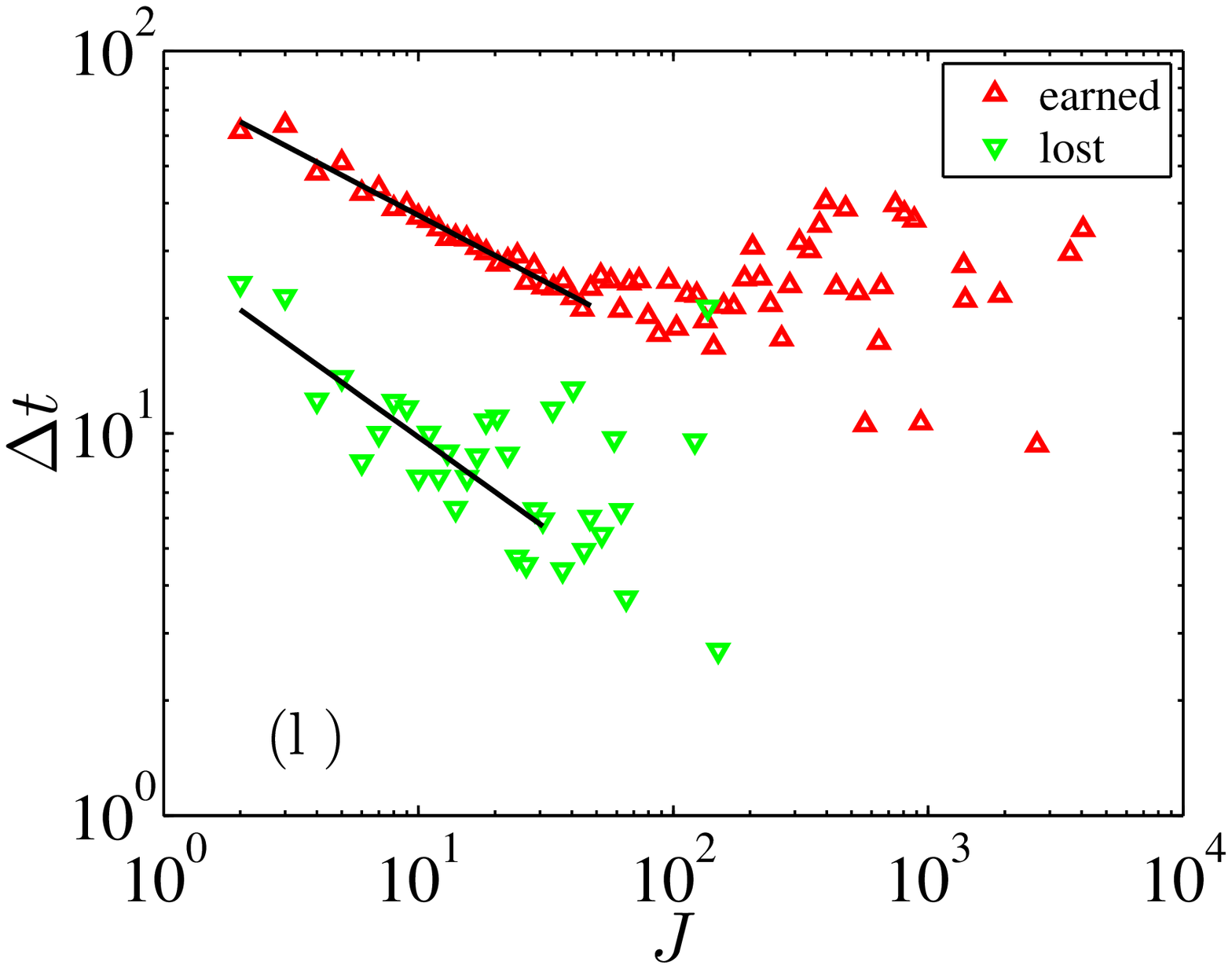}
  \caption{Power-law relationships between three variables. The first column (a,d,g,j) shows the dependence between the magnitude of the average return $R$ and the trading frequency $J$. The second column (b,e,h,k) shows the dependence between the magnitude of the average return $R$ and the holding time $\Delta t$. The third column (c,f,i,l) shows the dependence between the magnitude of the average return $R$ and the holding time $\Delta t$ and  the trading frequency $J$. The four rows are for A-share individuals, A-share institutions, B-share individuals, and B-share institutions, respectively.}
\label{Fig:PL}
\end{figure}

The winners or losers (individuals or institutions in a market) are sorted according to their trading frequencies. The averages of the returns and holding times of each group of investors are calculated. We plot the magnitude of the average returns for winners and for losers as a function of the trading frequency in double logarithmic coordinates in the first column of Fig.~\ref{Fig:PL} and observe a power law relationship
\begin{equation}
  R\sim J^{-\alpha},
  \label{Eq:PL:RJ}
\end{equation}
where $\alpha=0.31\pm0.01$ for A-share individual winners, $\alpha=0.38\pm0.01$ for A-share individual losers, $\alpha=0.20\pm0.04$  for A-share institutional winners, $\alpha=0.11\pm0.04$ for A-share institutional losers, $0.20\pm0.01$ for B-share individual winners, $0.39\pm0.02$ for B-share individual losers, $0.24\pm0.02$ for B-share institutional winners, and $0.46\pm0.09$ for B-share institutional losers.

Similarly, the magnitude of the average return $R$ scales with respect to the average holding time $\Delta{t}$ as a power law
\begin{equation}
  R\sim \Delta{t}^{\beta},
  \label{Eq:PL:RDt}
\end{equation}
and the average holding time $\Delta{t}$ also scales with respect to the average trading frequency $J$ as a power law
\begin{equation}
  \Delta{t}\sim J^{-\gamma},
  \label{Eq:PL:DtJ}
\end{equation}
as illustrated in the second and third columns of Fig.~\ref{Fig:PL}. The power laws are statistically more significant for individual investors than institutional investors because there are many more individuals (about 99.38\%) in the A-share market. The estimated exponents are listed in Table \ref{TB:PL:exponents}. For each power-law relationship, the exponents for winners and losers are approximately equal to each other, that is, 
\begin{equation}
  E_{\mathrm{winner,investor,market}} \approx E_{\rm{loser,investor,market}},
\end{equation}
where $E=\alpha$, $\beta$, or $\gamma$,``investor'' could be individuals or institutions, and ``market'' could be A-shares or B-shares.

Combining Eqs.~(\ref{Eq:PL:RJ}-\ref{Eq:PL:DtJ}), we obtain an equation relating the three power-law exponents
\begin{equation}
  \alpha=\beta\gamma,
  \label{Eq:PL:alpha:beta:gamma}
\end{equation}
which is validated in Table \ref{TB:PL:exponents}.

\begin{table}[htb]
 \centering
 \caption{{\bf{Estimated exponents of the power-law relationships}}}
 \label{TB:PL:exponents}
 {\tiny{
 \begin{tabular}{cccccccccccccccccccccccccccccccccccccccccccc}
  \hline\hline
  && \multicolumn{5}{c}{A-shares} && \multicolumn{5}{c}{B-shares}\\
   \cline{3-7}\cline{9-13}
  && \multicolumn{2}{c}{Individual} && \multicolumn{2}{c}{Institution} && \multicolumn{2}{c}{Individual} && \multicolumn{2}{c}{Institution}\\
   \cline{3-4}\cline{6-7}\cline{9-10}\cline{12-13}
   && Winner & Loser && Winner & Loser && Winner & Loser && Winner & Loser  \\ \hline
  $\alpha$   && $0.31\pm0.01$ & $0.38\pm0.01$ && $0.20\pm0.04$ & $0.11\pm0.04$ && $0.20\pm0.01$ & $0.39\pm0.02$ && $0.24\pm0.02$ & $0.46\pm0.09$\\
  $\beta$    && $1.37\pm0.10$ & $1.32\pm0.03$ && $0.39\pm0.22$ & $0.48\pm0.15$ && $0.94\pm0.04$ & $1.43\pm0.12$ && $0.79\pm0.04$ & $0.60\pm0.26$\\
  $\gamma$   && $0.20\pm0.01$ & $0.29\pm0.01$ && $0.07\pm0.03$ & $0.16\pm0.03$ && $0.22\pm0.01$ & $0.27\pm0.02$ && $0.35\pm0.01$ & $0.48\pm0.08$\\
  $\beta\gamma$  && $0.27\pm0.03$ & $0.38\pm0.02$ && $0.03\pm0.03$ & $0.08\pm0.04$ && $0.21\pm0.02$ & $0.39\pm0.06$ && $0.28\pm0.02$ & $0.29\pm0.17$\\
  \hline\hline
   \end{tabular}
   }}
\end{table}

\section*{Discussion}
\label{S1:Conclusion}

The original incentive of this study was to provide novel comparative
characterizations of financial markets that are based on the realized
performances of strategies implemented by investors.
The Shenzhen Stock Exchange of China offers a unique opportunity
to compare (i) a closed national market (A-shares) with an international market (B-Shares),
(ii) individuals and institutions and (iii) real investors to random strategies with respect
to timing that share otherwise all other characteristics.

The first robust result is that
more trading results in smaller net return due to trading frictions. This
is true for both individual and institutional investors in China's
B-share market. However, the net return of individual investors in the
A-share market is independent of the trading frequency, which is different from
other markets \cite{Odean-1999-AER,Barber-Odean-2000-JF,Barber-Odean-Zhu-2009-RFS,Barber-Lee-Liu-Odean-2009-RFS}.
For individual or institutional winners,
this result holds again. We unveiled quantitative laws
showing how the deterioration of performance scales with frequency and with holding period.
Naively, we could have expected that the performance is simply
inversely proportional to the trading frequency, if transaction costs was
the only contribution. But here, we find
non-trivial exponents, which reveal the complexity of the market price
structure as the investors strategically adapt their investments.
These results provide a new set of stylized facts that
characterize the structure of the price patterns. In other words,
the properties of the returns obtained by different investors provide a
kind of ``spectroscopy'' of the prices.

We also found that the return of real trading is significantly and robustly worse than random
trading. As a consequence, we can conclude that investors do try
to develop opportunistic strategies, but
zero intelligence strategies outperform them in stock trading. Certainly,
this conclusion does not deny the possibility that some investors do
perform better than random trading. Therefore, we can use the strategy
performance as a gauge or an instrument to characterize the market
structure, in addition to the its statistical properties often
referred to as the stylized facts. To the best of our knowledge, this idea is novel.
It reflects the natural consequence that the aggregation of strategies
make the stock market structure what it is, and vice-versa the later
influences and co-evolve with the ecology of strategies \cite{Farmer-2002-ICC}.
The strategies implemented by investors are not only probing the prices but also
influencing the prices so that they are both cameras and engines \cite{MacKenzie-2008}.
We believe that the study of the combined structure of both
strategies and prices will open a qualitatively new level of
understanding of financial and economic markets.


\section*{Acknowledgments}

WXZ and GHM acknowledge financial support from the National Natural Science Foundation of China under grant 11075054, the ``Shu Guang'' Project sponsored by the Shanghai Municipal Education Commission and the Shanghai Education Development Foundation under grant 2008SG29, and the Fundamental Research Funds for the Central Universities. WC acknowledges financial support from the National Natural Science Foundation of China under grant 11075054. DS acknowledges financial support from the ETH Competence Center ``Coping with Crises in Complex Socio-Economic Systems'' (CCSS) through ETH Research Grant CH1-01-08-2 and ETH Zurich Foundation and the ETH Research Grant ETH-31 10-3 ``Testing the predictability of financial bubbles and of systemic instabilities''.

\bibliography{E:/Papers/Auxiliary/Bibliography}

\begin{thebibliography}{10}
\providecommand{\url}[1]{\texttt{#1}}
\providecommand{\urlprefix}{URL }
\expandafter\ifx\csname urlstyle\endcsname\relax
  \providecommand{\doi}[1]{doi:\discretionary{}{}{}#1}\else
  \providecommand{\doi}{doi:\discretionary{}{}{}\begingroup
  \urlstyle{rm}\Url}\fi
\providecommand{\bibAnnoteFile}[1]{%
  \IfFileExists{#1}{\begin{quotation}\noindent\textsc{Key:} #1\\
  \textsc{Annotation:}\ \input{#1}\end{quotation}}{}}
\providecommand{\bibAnnote}[2]{%
  \begin{quotation}\noindent\textsc{Key:} #1\\
  \textsc{Annotation:}\ #2\end{quotation}}
\providecommand{\eprint}[2][]{\url{#2}}

\bibitem{Dobzhansky-1973-ABT}
Dobzhansky T (1973) {Nothing in biology makes sense except in the light of
  evolution}.
\newblock Amer Bio Teacher 35: 125-129.
\bibAnnoteFile{Dobzhansky-1973-ABT}

\bibitem{Haldane-May-2011-Nature}
Haldane AG, May RM (2011) {Systemic risk in banking ecosystems}.
\newblock Nature 469: 351-355.
\bibAnnoteFile{Haldane-May-2011-Nature}

\bibitem{Johnson-2011-Nature}
Johnson N (2011) {Proposing policy by analogy is risky}.
\newblock Nature 469: 302.
\bibAnnoteFile{Johnson-2011-Nature}

\bibitem{Lux-2011-Nature}
Lux T (2011) {Network theory is sorely required}.
\newblock Nature 469: 303.
\bibAnnoteFile{Lux-2011-Nature}

\bibitem{Cont-2001-QF}
Cont R (2001) {Empirical properties of asset returns: Stylized facts and
  statistical issues}.
\newblock Quant Financ 1: 223-236.
\bibAnnoteFile{Cont-2001-QF}

\bibitem{Sornette-Woodard-2009-XXX}
Sornette D, Woordard R (2009) {Real Estate bubbles, Derivative Bubbles, and the
  Financial and Economic Crisis}.
\newblock To appear in the Proceedings of APFA7 (Applications of Physics in
  Financial Analysis), ``New Approaches to the Analysis of Large-Scale Business
  and Economic Data,'' Misako Takayasu, Tsutomu Watanabe and Hideki Takayasu,
  eds., Springer (2010)
  (\url{http://papers.ssrn.com/sol3/papers.cfm?abstract_id=1407608}).
\bibAnnoteFile{Sornette-Woodard-2009-XXX}

\bibitem{Bernanke-2004}
Bernanke BS (2004) {The Great Moderation}.
\newblock Remarks by Governor Ben S. Bernanke at the meetings of the Eastern
  Economic Association, Washington, DC, February 20, 2004.
\bibAnnoteFile{Bernanke-2004}

\bibitem{LeBaron-Arthur-Palmer-1999-JEDC}
LeBaron B, Arthur WB, Palmer R (1999) {Time series properties of an artificial
  stock market}.
\newblock J Econ Dyn Control 23: 1487-1516.
\bibAnnoteFile{LeBaron-Arthur-Palmer-1999-JEDC}

\bibitem{Hommes-2001-QF}
Hommes CH (2001) {Financial markets as nonlinear adaptive evolutionary
  systems}.
\newblock Quant Financ 1: 149-167.
\bibAnnoteFile{Hommes-2001-QF}

\bibitem{Hommes-2002-PNAS}
Hommes CH (2002) {Modeling the stylized facts in finance through simple
  nonlinear adaptive systems}.
\newblock Proc Natl Acad Sci USA 99: 7221-7228.
\bibAnnoteFile{Hommes-2002-PNAS}

\bibitem{Farmer-2002-ICC}
Farmer JD (2002) {Market force, ecology and evolution}.
\newblock Industrial and Corporate Change 11: 895-953.
\bibAnnoteFile{Farmer-2002-ICC}

\bibitem{Ehrentreich-2006-JEBO}
Ehrentreich N (2006) {Technical trading in the Santa Fe Institute Artificial
  Stock Market revisited}.
\newblock J Econ Behav Org 61: 599-616.
\bibAnnoteFile{Ehrentreich-2006-JEBO}

\bibitem{Hommes-Wagener-2009}
Hommes C, Wagener F (2009) {Complex Evolutionary Systems in Behavioral
  Finance}.
\newblock Amsterdam, The Netherlands: North-Holland, Elsevier, Inc.
\newblock Chapter 4 of Handbook of Financial Markets: Dynamics and Evolution,
  pp. 217-276.
\bibAnnoteFile{Hommes-Wagener-2009}

\bibitem{Lo-2004-JPM}
Lo AW (2004) {The adaptive markets hypothesis}.
\newblock J Portfolio Management 30: 15-29.
\bibAnnoteFile{Lo-2004-JPM}

\bibitem{Gode-Sunder-1993-JPE}
Gode DK, Sunder S (1993) {Allocative efficiency of markets with
  zero-intelligence traders: Market as a partial substitute for individual
  rationality}.
\newblock J Polit Econ 101: 119-137.
\bibAnnoteFile{Gode-Sunder-1993-JPE}

\bibitem{Othman-2008}
Othman A (2008) {Zero-Intelligence Agents in Prediction Markets}.
\newblock In: Padgham M Parkes, Parsons, editors, Proc. of 7th Int. Conf. on
  Autonomous Agents and Multiagent Systems. Estoril, Portugal: AAMAS 2008, pp.
  879-886.
\bibAnnoteFile{Othman-2008}

\bibitem{Farmer-Patelli-Zovko-2005-PNAS}
Farmer JD, Patelli P, Zovko II (2005) {The predictive power of zero
  intelligence in financial markets}.
\newblock Proc Natl Acad Sci USA 102: 2254-2259.
\bibAnnoteFile{Farmer-Patelli-Zovko-2005-PNAS}

\bibitem{Malkiel-2011}
Malkiel BG (2011) {A Random Walk Down Wall Street: The Time-Tested Strategy for
  Successful Investing}.
\newblock New York: W. W. Norton \& Company.
\bibAnnoteFile{Malkiel-2011}

\bibitem{Barras-Scaillet-Wermers-2010-JF}
Barras L, Scaillet O, Wermers R (2010) {False discoveries in mutual fund
  performance: Measuring luck in estimated alphas}.
\newblock J Financ 65: 179-216.
\bibAnnoteFile{Barras-Scaillet-Wermers-2010-JF}

\bibitem{Fama-French-2010-JF}
Fama E, French K (2010) {Luck versus skill in the cross-section of mutual fund
  returns}.
\newblock J Financ 65: 1915-1947.
\bibAnnoteFile{Fama-French-2010-JF}

\bibitem{Kosowski-Timmermann-Wermers-White-2006-JF}
Kosowski R, Timmermann A, Wermers R, White H (2006) {Can mutual fund ``stars''
  really pick stocks? New evidence from a bootstrap analysis}.
\newblock J Financ 61: 2551-2595.
\bibAnnoteFile{Kosowski-Timmermann-Wermers-White-2006-JF}

\bibitem{Satinover-Sornette-2007-EPJB}
Satinover JB, Sornette D (2007) {¡°Illusion of control¡± in time-horizon
  minority and parrondo games}.
\newblock Eur Phys J B 60: 369-384.
\bibAnnoteFile{Satinover-Sornette-2007-EPJB}

\bibitem{Gu-Chen-Zhou-2007-EPJB}
Gu GF, Chen W, Zhou WX (2007) {Quantifying bid-ask spreads in the Chinese stock
  market using limit-order book data: Intraday pattern, probability
  distribution, long memory, and multifractal nature}.
\newblock Eur Phys J B 57: 81-87.
\bibAnnoteFile{Gu-Chen-Zhou-2007-EPJB}

\bibitem{DeMiguel-Garlappi-Uppal-2009-RFS}
DeMiguel V, Garlappi L, Uppal R (2009) {Optimal versus naive diversification:
  How enefficient is the 1/N portfolio strategy}.
\newblock Rev Financ Stud 53: 1915-1953.
\bibAnnoteFile{DeMiguel-Garlappi-Uppal-2009-RFS}

\bibitem{Parrondo-1996}
Parrondo JMR (1996) {Efficiency of Brownian motors}.
\newblock Workshop of the EEC HC\&M Network on Complexity and Chaos.
\bibAnnoteFile{Parrondo-1996}

\bibitem{Harmer-Abbott-1999-SS}
Harmer GP, Abbott D (1999) {Parrondo's paradox}.
\newblock Statist Sci 14: 206-213.
\bibAnnoteFile{Harmer-Abbott-1999-SS}

\bibitem{Harmer-Abbott-1999-Nature}
Harmer GP, Abbott D (1999) {Losing strategies can win by Parrondo's Paradox}.
\newblock Nature 402: 864.
\bibAnnoteFile{Harmer-Abbott-1999-Nature}

\bibitem{Harmer-Abbott-Taylor-Parrondo-2000-UPNF}
Harmer GP, Abbott D, Taylor PG, Parrondo JMR (2000) {Parrondo's paradoxical
  games and the discrete Brownian ratchet}.
\newblock Unsolved Problems of Noise and Fluctuations 511: 189-200.
\bibAnnoteFile{Harmer-Abbott-Taylor-Parrondo-2000-UPNF}

\bibitem{Satinover-Sornette-2007-PA}
Satinover JB, Sornette D (2007) {Illusion of control in a Brownian game}.
\newblock Physica A 386: 339-344.
\bibAnnoteFile{Satinover-Sornette-2007-PA}

\bibitem{Satinover-Sornette-2009-EPJB}
Satinover JB, Sornette D (2009) {Illusory versus genuine control in agent-based
  games}.
\newblock Eur Phys J B 67: 357-367.
\bibAnnoteFile{Satinover-Sornette-2009-EPJB}

\bibitem{Barber-Odean-Zhu-2009-RFS}
Barber BM, Odean T, Zhu N (2009) {Do retail trades move markets?}
\newblock Rev Financ Stud 22: 151-186.
\bibAnnoteFile{Barber-Odean-Zhu-2009-RFS}

\bibitem{Torngren-Montgomery-2004-JBF}
Torngren G, Montgomery H (2004) {Worse than chance? Performance and confidence
  among professionals and laypeople in the stock market}.
\newblock J Behav Financ 5: 148-153.
\bibAnnoteFile{Torngren-Montgomery-2004-JBF}

\bibitem{Ljungqvist-Malloy-Marston-2009-JF}
Ljungqvist A, Malloy C, Marston F (2009) {Rewriting history}.
\newblock J Financ 64: 1935-1960.
\bibAnnoteFile{Ljungqvist-Malloy-Marston-2009-JF}

\bibitem{Daniel-Sornette-Woehrmann-2009-JPM}
Daniel G, Sornette D, Woehrmann P (2009) {Look-ahead benchmark bias in
  portfolio performance evaluation}.
\newblock J Portfolio Management 36: 121-130.
\bibAnnoteFile{Daniel-Sornette-Woehrmann-2009-JPM}

\bibitem{Odean-1998b-JF}
Odean T (1998) {Volume, volatility, price and profit when all traders are above
  average}.
\newblock J Financ 53: 1887-1934.
\bibAnnoteFile{Odean-1998b-JF}

\bibitem{Gervais-Odean-2001-RFS}
Gervais S, Odean T (2001) {Learning to be overconfident}.
\newblock Rev Financ Stud 14: 1-27.
\bibAnnoteFile{Gervais-Odean-2001-RFS}

\bibitem{Statman-Thorley-Vorkink-2006-RFS}
Statman M, Thorley S, Vorkink K (2006) {Investor overconfidence and trading
  volume}.
\newblock Rev Financ Stud 19: 1531-1565.
\bibAnnoteFile{Statman-Thorley-Vorkink-2006-RFS}

\bibitem{Glaser-Weber-2007-GRIR}
Glaser M, Weber M (2007) {Overconfidence and trading volume}.
\newblock Geneva Risk Insur Rev 32: 1-36.
\bibAnnoteFile{Glaser-Weber-2007-GRIR}

\bibitem{Glaser-Weber-2009-JFinM}
Glaser M, Weber M (2009) {Which past returns affect trading volume?}
\newblock J Financ Markets 12: 1-31.
\bibAnnoteFile{Glaser-Weber-2009-JFinM}

\bibitem{Deaves-Luders-Luo-2009-RF}
Deaves R, L{\"u}ders E, Luo GY (2009) {An experimental test of the impact of
  overconfidence and gender on trading activity}.
\newblock Rev Financ 13: 555-575.
\bibAnnoteFile{Deaves-Luders-Luo-2009-RF}

\bibitem{Odean-1999-AER}
Odean T (1999) {Do investors trade too much?}
\newblock Amer Econ Rev 89: 1279-1298.
\bibAnnoteFile{Odean-1999-AER}

\bibitem{Barber-Odean-2000-JF}
Barber BM, Odean T (2000) {Trading is hazardous to your wealth: The common
  stock investment performance of individual investors}.
\newblock J Financ 55: 773-806.
\bibAnnoteFile{Barber-Odean-2000-JF}

\bibitem{Barber-Lee-Liu-Odean-2009-RFS}
Barber BM, Lee Y, Liu Y, Odean T (2009) {Just how much do individual investors
  lose by trading?}
\newblock Rev Financ Stud 22: 609-632.
\bibAnnoteFile{Barber-Lee-Liu-Odean-2009-RFS}

\bibitem{Sornette-Zhou-2005-QF}
Sornette D, Zhou WX (2005) {Non-parametric determination of real-time lag
  structure between two time series: The ``optimal thermal causal path''
  method}.
\newblock Quant Financ 5: 577-591.
\bibAnnoteFile{Sornette-Zhou-2005-QF}

\bibitem{MacKenzie-2008}
MacKenzie D (2008) {An Engine, Not a Camera: How Financial Models Shape
  Markets}, volume~1.
\newblock London: The MIT Press.
\bibAnnoteFile{MacKenzie-2008}

\end{thebibliography}

\end{document}